\documentclass[useAMS,usenatbib,usegraphicx]{mn2e}
\usepackage{amsmath}
\usepackage[toc,page]{appendix}

	\usepackage{pslatex}
	\usepackage{amsmath}

\title[ Evolution of a Viscous Protoplanetary Disk]{Evolution of a Viscous Protoplanetary Disk with Convectively Unstable Regions. 
II. Accretion Regimes and Long-Term Dynamics}
\author[Maksimova et al.]{L.A. Maksimova, Ya.N. Pavlyuchenkov, A.V. Tutukov
\\
Institute of Astronomy, Russian Academy of Sciences, Moscow, Russia
}

\begin{document}
\date{Received April 29, 2020. Accepted  June 30, 2020.  \\
lomara.maksimova@gmail.com}
\pagerange{815--826} \pubyear{2020}
\volume{64}
\maketitle

%
\begin{abstract}

In this article, we proceed to study convection as a possible factor of episodic accretion in protoplanetary disks. 
Within the model presented in Article~I, the accretion history is analyzed at different rates and areas of matter inflow from the envelope onto the disk. It is shown that the burst-like regime occurs in a wide range of parameters. The long-term evolution of the disk is also modeled, including the decreasing-with-time matter inflow from the envelope. 
It is demonstrated that the disk becomes convectively unstable and maintains burst-like accretion onto the star for several million years. Meanwhile, the
instability expands to an area of several tens of astronomical units and gradually decreases with time. 
It is also shown that at early stages in the disk evolution, conditions arise for gravitational instability in the outer parts of the disk and for dust evaporation in the convectively unstable inner regions. The general conclusion of the study is that convection can serve as one of the mechanisms of episodic accretion in protostellar disks, but this conclusion needs to be verified using more consistent hydrodynamic models.

DOI: 10.1134/S1063772920110050
\end{abstract}  


\def\gtrsim{\mathrel{\hbox{\rlap{\hbox{\lower5pt\hbox{$\sim$}}}\hbox{$>$}}}}

\section{INTRODUCTION}

The formation and evolution of protoplanetary
disks (PDs) around young stars is one of the most intriguing
topics in astrophysics. In the earliest evolution stages,
many young stellar systems show signs of episodic
1 accretion; these objects are known as FUors and EXors
(see, e.g.,~\citep{1996ARA&A..34..207H,2014prpl.conf..387A}). There are theoretical agruments to believe that all PDs go
through a phase of episodic accretion at initial stages
in their evolution, which explains the variable luminosity of young stellar objects~\citep{1998apsf.book.....H}. However, the physical mechanisms of this variability remain debatable.
The variability problem is closely related to the more
general question about the mechanisms of angular
momentum transfer in accretion disks. Vigorous discussions have taken place regarding the possible
mechanisms that not only ensure the transfer of angular momentum in disks but also cause the nonsteady
accretion pattern and involve such factors as gravitational,
magnetorotational, and thermal instabilities (see~\citep{2014prpl.conf..387A}).
Thus, the nonsteady pattern of accretion caused by
gravitational instability is associated with the clump
formation and their falling onto the star, leading to
luminosity bursts (see, e.g.,~\citep{2006ApJ...650..956V}. The burst-like pattern caused by magnetorotational instability is due to
the positive feedback link of this instability to the ionization degree of matter~\citep{2009ApJ...701..620Z}. Thermal instability is
caused by an increase in gas opacity with increasing
temperature in partially ionized gas~\citep{1999ApJ...518..833K}.

Convection is also considered as one of the mechanisms responsible for the turbulization of matter in
accretion disks (see, e.g.,~\citep{1980MNRAS.191...37L,
2015MNRAS.451.3995S,2017MNRAS.464..410M,2018MNRAS.480.4797H}). In our previous
work~\citep{2020ARep...64....1P}, which will be referred to below as Article~I,
we demonstrated that convective instability can also
lead to a nonsteady pattern of accretion. Within the
viscous disk model, we showed that in the presence of
background viscosity, convection is a process with a positive feedback due to an increase in the opacity of gas-and-dust medium with increasing temperature and can therefore be responsible for bursts and episodic accretion in PDs. However, we neither investigated how the model parameters influence this conclusion nor studied the nature of episodic accretion for different rates of matter inflow from the envelope.

The aim of our work is a more detailed study of the
nonsteady accretion regime that was obtained in Article~I. The present article is organized as follows. Section~\ref{sec_descr} gives a brief description of the model. Section~\ref{sec_param}
examines the effect of the model parameters on the
accretion pattern within the simplest description of
matter inflow from the envelope, which is used in Article~I. Section~\ref{sec_evol} studies the diskТs long-term evolution
with a more realistic description of matter accretion
from the envelope. The conclusions section summarizes the results of this study.

\section{MODEL DESCRIPTION}
\label{sec_descr}
In this work, we use the PD evolution model from
Article~I. Here, we outline only the key elements of
this model. We consider an axially symmetric, geometrically thin viscous Keplerian disk without a radial
pressure gradient. The evolution of the surface density
of the disk is modeled using the Pringle equation~\citep{1981ARA&A..19..137P},
considering the accretion of matter from the envelope
onto the disk:
\begin{equation}
\frac{\partial \Sigma}{\partial t} =
\frac{3}{R}\frac{\partial }{\partial R}\left[\sqrt{R}\frac{\partial}{\partial R}
\left(\nu\sqrt{R}\Sigma\right)\right] + W(R,t),
\label{eq1}
\end{equation}
where $\Sigma(R,t)$ is the surface viscosity; $R$ is the distance
to the star; $t$ is time; $\nu(R,t)$ is the kinematic viscosity
coefficient; and $W(R,t)$ is the matter inflow from the
envelope under the assumption that the specific angular momentum of the settling matter coincides with
that in the disk. To identify convectively unstable
regions in the disk, radial evolution is modelled in parallel with reconstructing the diskТs vertical structure.
In the approximation of a hydrostatic-equilibrium
disk, the density and temperature distributions are calculated in the polar direction. The main factor governing the disk evolution within this model is the dependence of the viscosity coefficient $\nu(R,t)$ on the radius.
The model assumes that the viscosity coefficient is a
sum of the background and convective viscosity (the
idea of convection-related additional viscosity in circumstellar disks is discussed in more detail in~\citep{2003AN,2014ApJ...787....1H,2016MNRAS.462.3710C}):
\begin{equation}
\nu(R,t)=\nu_{\rm bg}(R)+\tilde{\nu}_{\rm c}(R,t).
\label{eq_nu_all}
\end{equation}
The background viscosity is associated with some
nonconvective mechanism of angular momentum
transfer (e.g., with magnetorotational instability~\citep{1991ApJ...376..214B,1991ApJ...376..223H}) and ensures continuous gas accretion. This viscosity is given as\begin{equation}
\nu_{\rm bg}(R) = \nu_{0}\left(\frac{R}{1~{\rm AU}}\right)^{\beta},
\label{eq_nu}
\end{equation}
where the parameter $\beta=1$ is chosen to reproduce the density distribution law in the observed PDs~\citep{2011ARA&A..49...67W}. 
The source of convection in this model is the heat release due to viscous dissipation of gas:
\begin{equation}
\Gamma_{\rm vis} = \frac{9}{4} \frac{GM_{*}}{R^3}\nu \Sigma,
\label{eq_nuheat}
\end{equation}
where $\Gamma_{\rm vis}$ is the rate of viscous dissipation per unit
area of the disk at a given radius and $M_{*}$ is the mass of the star.

The convective viscosity $\nu_{\rm c}$ is nonzero in convectively unstable regions and is calculated as
\begin{equation}
\nu_{\rm c} = \gamma\, H V_{\rm c},
\label{eq_nuconv}
\end{equation}
where $\gamma$ is the mass fraction of matter in a convectively
unstable region at a given radius, as calculated from
the procedure of reconstructing the diskТs vertical
structure; $H$ is the local disk height; and $V_{\rm c}$ is the
characteristic velocity of convective elements. Hereafter, the calculated
convective viscosity $\nu_{\rm c}$ is smoothed in the radial direction on the scale $H$, forming $\tilde{\nu}_{\rm c}$ in \eqref{eq_nu_all}. In Article~I, we
derived the characteristic velocity $V_{\rm c}$ under the
assumption that all the energy released as a result of
viscous dissipation transforms into the kinetic energy
of its convective motion, which is the upper estimate
for this value. In this work, we introduce a coefficient $\eta \le
1$, which characterizes the transition efficiency of
the thermal energy into convective motion:
\begin{equation}
\eta\Gamma_{\rm vis} = \frac{\rho_{0} V_{\rm c}^2}{2} V_{\rm c},
\label{eq_gammavis}
\end{equation}
where $\rho_{0}$ is the equatorial density.

When reconstructing the diskТs vertical structure,
we consider the heating by radiation from the central
object along with the viscous gas dissipation. The
luminosity of the central source consists of the photospheric
luminosity of the star (which is assumed to be equal to
that of the Sun) and the accretion luminosity, which is
calculated by the formula
\begin{equation}
L_{acc}=\dfrac{1}{2}\dfrac{GM_{*}\dot{M}_{*}}{R_{*}},
\label{acc_lum}
\end{equation}
where $R_{*}$ is the stellar radius and $\dot{M}_{*}$ is the rate of accretion from the disk to the star.

When solving Eq.~\eqref{eq1}, we use a fixed value of surface density on the inner (0.2~AU) and outer (200~AU)
boundaries of the model disk. The density values at the
boundaries were chosen comparatively small, which
ensures free outflow of matter from the region under
consideration. As noted in Article~I, more complex
boundary conditions require separate study. The
importance of the inner boundary condition in PD
modeling is the central topic of research, e.g., in~\citep{2019A&A...627A.154V}.

In this work, we study the effects of the four model
parameters on the manifestation of the episodic accretion pattern: (1) the accretion rate from the envelope
onto the disk, $\dot{M}$; (2) the accumulation region, onto which
matter from the envelope accretes, $R_\text{ring}$; (3) the back-
ground viscosity $\nu_{0}$; and (4) the convection efficiency
coefficient $\eta$. In Article~I, we set a constant inflow of
gas from the envelope onto the disk into a ring between
10 and 20~AU with an accretion rate of $10^{-7}~M_{\odot}$/yr, and we also used the coefficients $\nu_{0}=10^{15}$~cm$^2$/s and $\eta=1$. This model will be referred to below as the \textit{basic model}. In this work, we consider models with accretion
rates differing from the basic one by two orders of
magnitude in both directions, which corresponds to
the spread of the accretion rates observed in PDs~\citep{2006A&A...452..245N,2014MNRAS.439..256E}. Along with an accumulation region of 10Ц20~AU, we also
consider a case where gas accretes onto the disk into a ring between 1 and 2~AU, which is more typical of initial stages in the evolution of PDs.

Another important parameter of the model is background viscosity, which predetermines the lifetime and disk mass. In the basic model, we use $\nu_{0}=10^{15}$~cm$^2$/s, which corresponds to a relatively high, turbulent viscosity. 
Based on the estimates from Article~I,
the corresponding alpha parameter of turbulent viscosity~\citep{1972AZh....49..921S,1973A&A....24..337S} is $\alpha=0.1$. In the present article, we also
consider a model with background viscosity an order
of magnitude smaller than the basic one; this smaller
background viscosity is closer to the observed estimates for evolved PDs. The convection efficiency
coefficient $\eta=1$, taken in the basic model, intentionally overestimates the transition of thermal energy into
convective energy since it does not account for that a part of the energy must be
transferred by radiation. The question arises: Will the
bursts caused by convection disappear if a considerable part of the released energy is transferred by radiation? To answer this question, we examined a model
with $\eta=0.1$. Table~\ref{tab_models} presents the parameters of the
models, and Section~\ref{sec_param} describes the modeling results.

In addition to these models, we considered a model
for studying the long-term evolution of the disk with a
more realistic accretion rate and accumulation region, the
parameters of which change with time. The function $W(R,t)$, which corresponds to the latter case, and the
results are described in Section~\ref{sec_evol}.

\begin{table}[]
\begin{center}
\caption{Parameters of the models under consideration}
\bigskip
\begin{tabular}{c|c|c|c|c}
\hline
Model & $\dot{M}$, & $R_\text{ring}$, & $\eta$ &  $\nu_{0}/10^{15}$, \\
\# & M$_{\odot}$/yr & AU &     ---            & ~cm$^2$/s \\
\hline
1  & $10^{-9}$ & $10-20$  & 1 & 1 \\
2  & $10^{-7}$ & $10-20$  & 1 & 1 \\
3  & $10^{-5}$ & $10-20$  & 1 & 1 \\
\hline
4  & $10^{-9}$ & $1-2$  & 1 & 1 \\
5  & $10^{-7}$ & $1-2$  & 1 & 1 \\
6  & $10^{-5}$ & $1-2$  & 1 & 1 \\
\hline
7  & $10^{-7}$ & $10-20$  & 1 & 0.1 \\
8  & $10^{-7}$ & $10-20$  & $\>$ 0.1 $\>$ & 1 \\
\hline
\end{tabular}
\label{tab_models}
\end{center}
\end{table}

\section{DISK ACCRETION REGIMES}
\label{sec_param}
This section presents the results for the models with
a constant (in time and space) inflow of matter from
the envelope onto the disk. The corresponding rate of
matter inflow inside the accumulation region is
\begin{equation}
W(R,t) = \dfrac{\dot{M}}{\pi(R_\text{ring,2}^2-R_\text{ring,1}^2)},
\end{equation}
where $\dot{M}$ is the integral rate of inflow from the envelope onto the disk; $R_\text{ring,1}$ and $R_\text{ring,2}$ are the inner and
outer radii of the accumulation region. We conducted the
modeling until the onset of the bursts, if any
occurred, and studied the characteristics of the appearing bursts. Thus, we
ignored the subsequent evolution of the disk. In the
calculations, we assumed the mass ($M_{*}=1\, M_\odot$), radius ($R_{*}=1\, R_\odot$), and luminosity ($L_{*}=1\, L_\odot$) of the central star to be constant. Meanwhile, we considered
the variability of the integral luminosity of the central
object due to the matter accretion from the disk onto
the star (see formula~\eqref{acc_lum}).

\begin{figure*}
\hfill
\includegraphics[width=1\columnwidth]{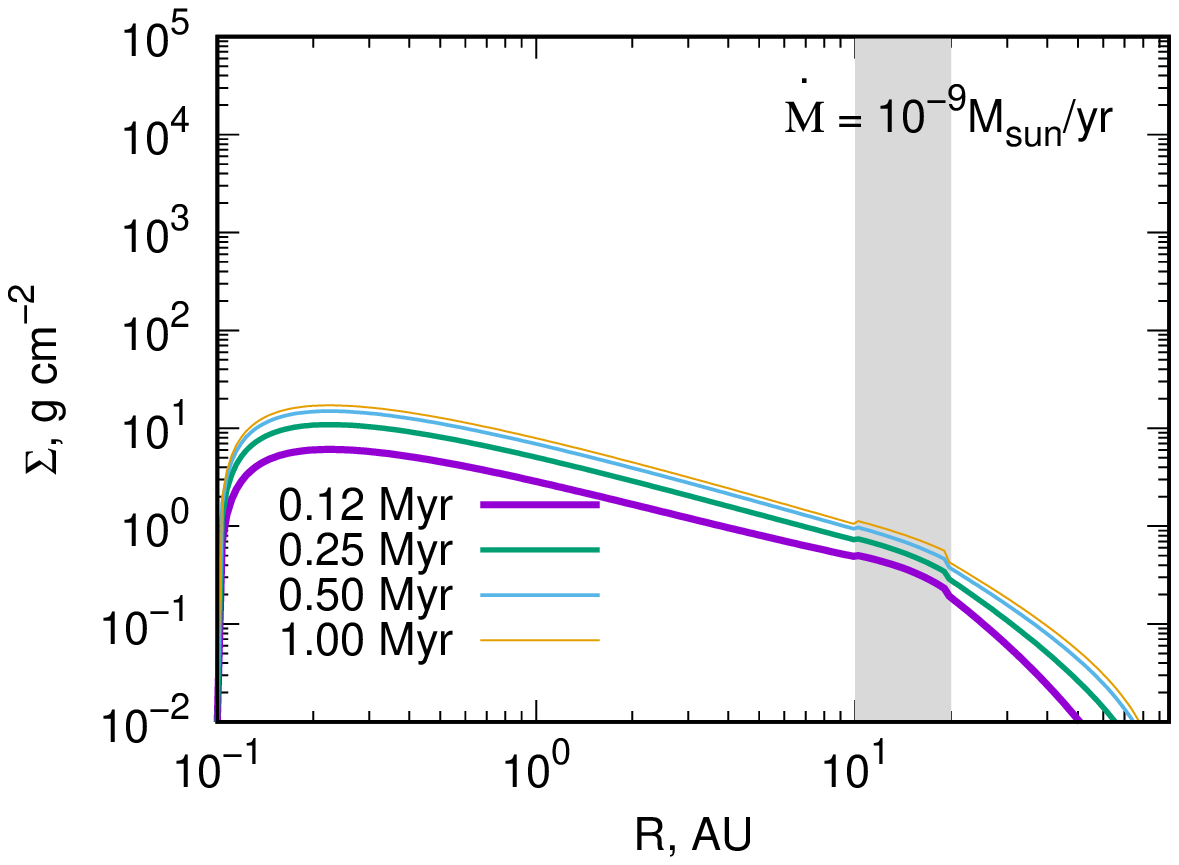}
\includegraphics[width=1\columnwidth]{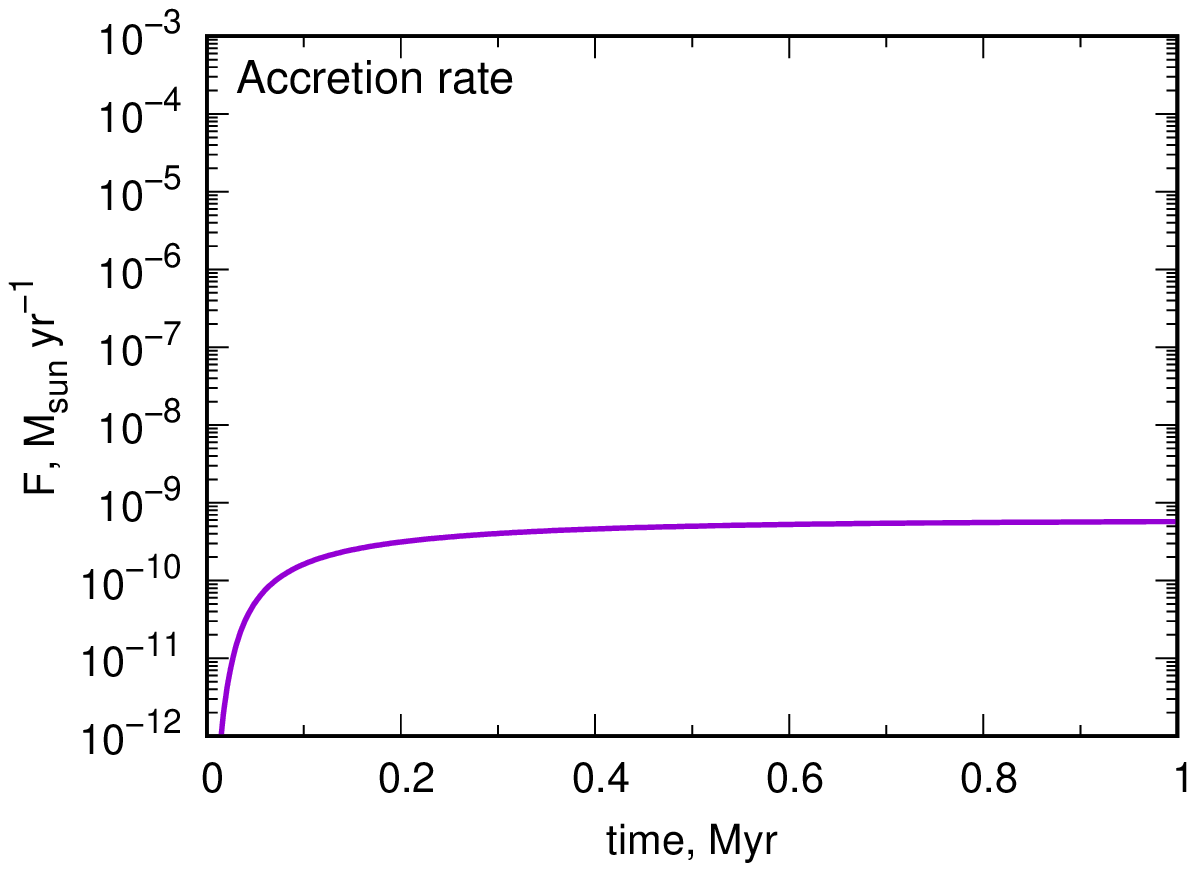}\\
\hfill
\includegraphics[width=1\columnwidth]{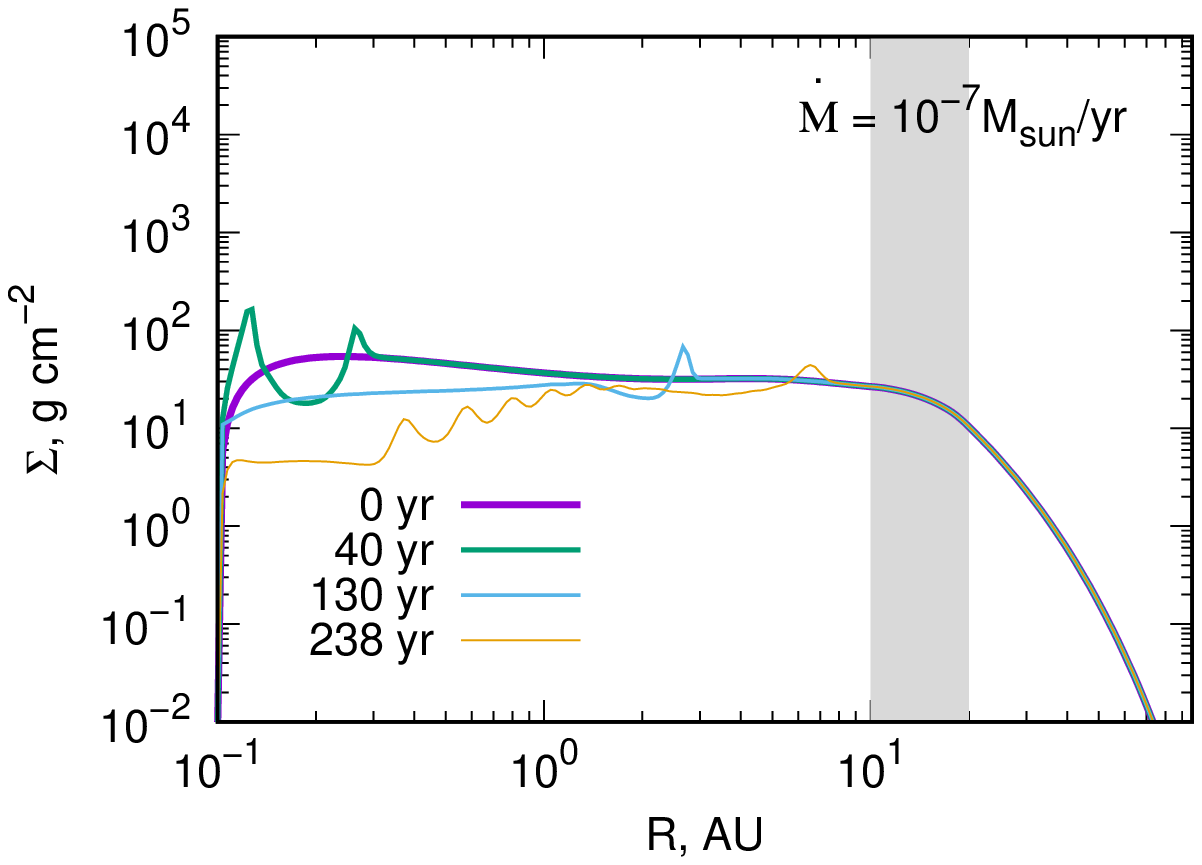}
\includegraphics[width=1\columnwidth]{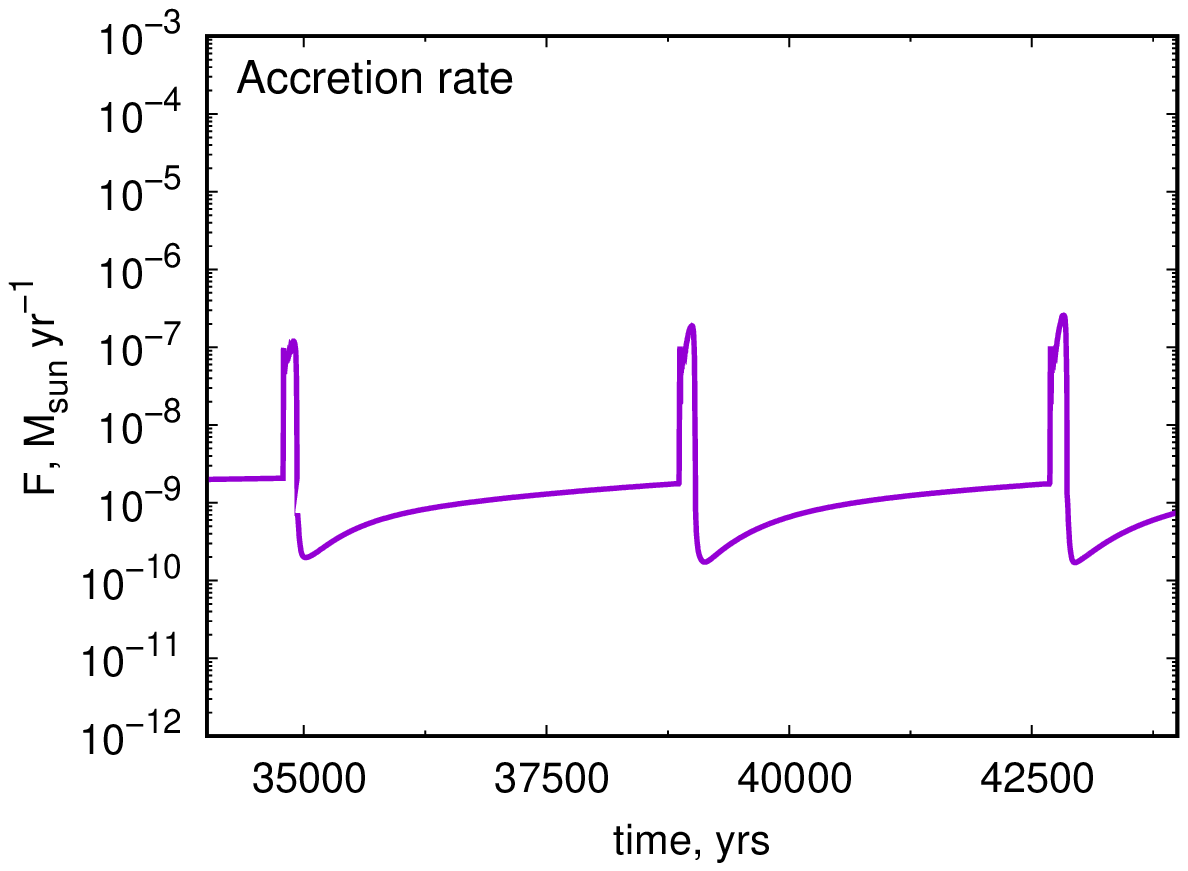}\\
\hfill
\includegraphics[width=1\columnwidth]{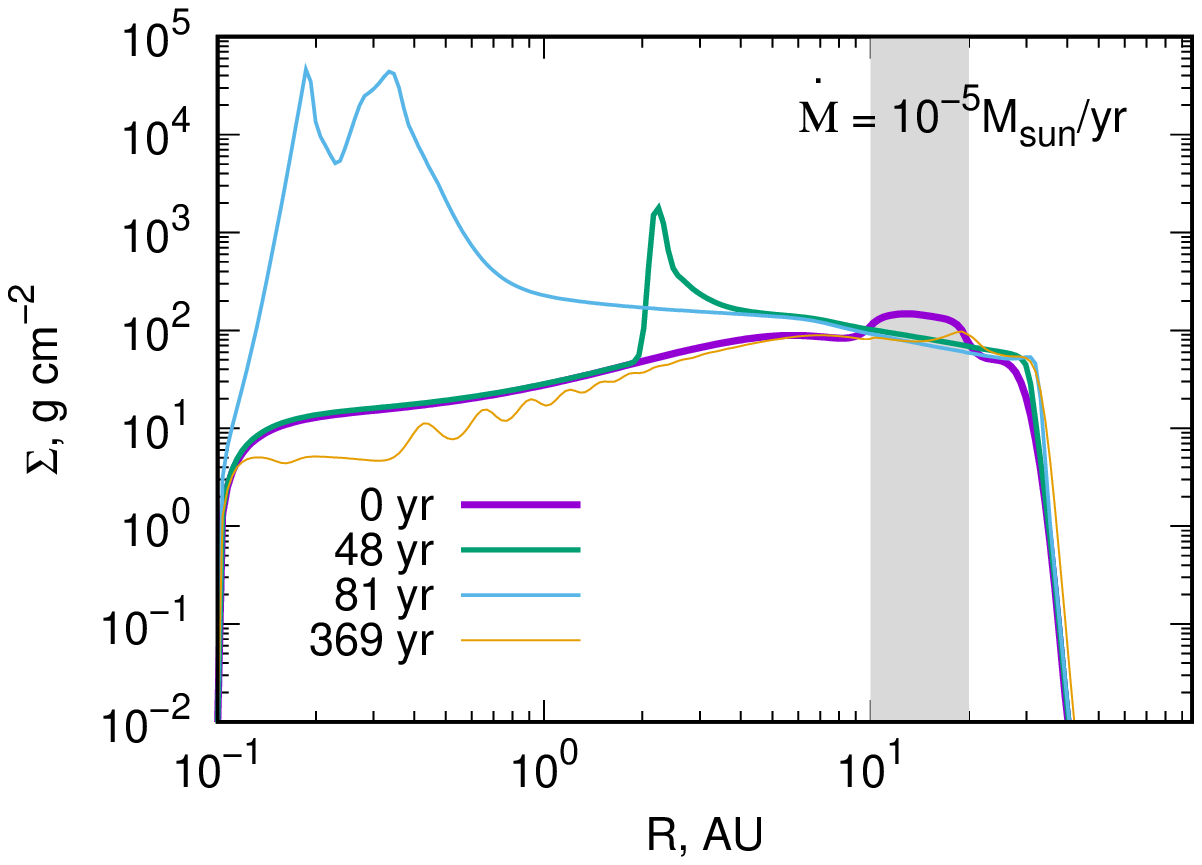}
\includegraphics[width=1\columnwidth]{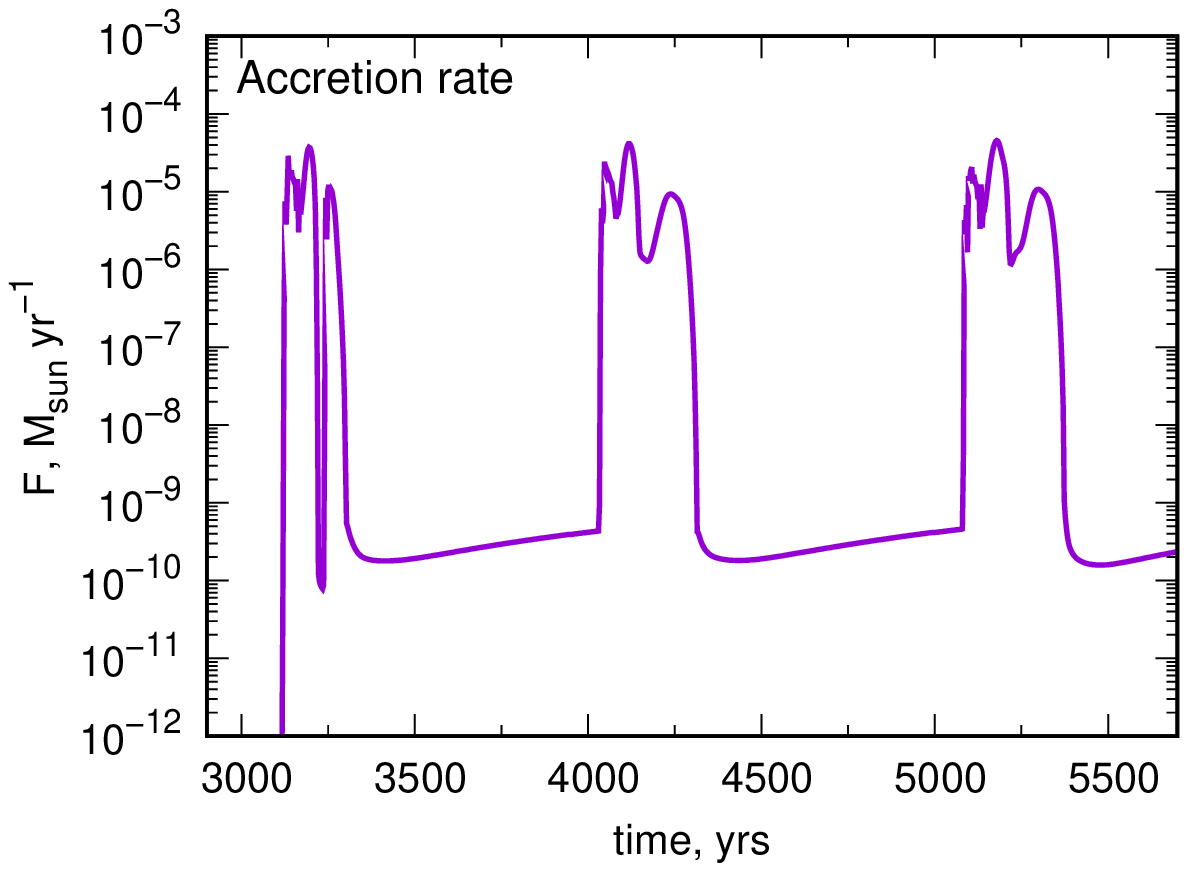}\\
\vspace{1cm}
\caption{
Simulation results for disk evolution in the models with an outer accumulation region (the top, middle, and bottom panels for Models~1, 2, and 3, respectively). Left: the radial surface-density distributions illustrating the development of an accretion burst. Time is counted from the end of the previous accretion burst. The vertical band shows the region of gas accretion from the envelope. Right: the accretion rate of matter from the disk onto the star.}
\label{evolmdot}
\end{figure*}

\subsection{Models with an Outer Accumulation Region}
We consider modeling results for cases with an
outer accumulation region ($R_\text{ring}=10-20$~AU), which differ
in the accretion rates from the envelope (Models~1, 2,
and 3) under the fixed $\nu_{0}=10^{15}$~cm$^2$/s and $\eta=1$. The
evolution of the surface density distributions and the
rates of gas accretion from the disk onto the star for
these models are shown in Fig.~\ref{evolmdot}. Model~1, with $10^{-9}$~M/yr (Fig.~\ref{evolmdot}, top panel), sets a quasi-stationary
regime, as seen from the time dependence of $\dot{M}$. The
density distributions for different times have a smooth
shape with small features near the accumulation region and
the diskТs inner boundary, which are due to the
boundary conditions. Given this relatively low accretion rate, convectively unstable regions do not appear
in the disk, and its evolution is fully defined by the
background viscosity. The accretion rate from the disk
onto the star approaches a stationarity with a value close
to the rate of matter inflow from the envelope.

When the rate of gas inflow onto the disk increases
to $10^{-7}$~M$_{\odot}$/yr (Model~2), this creates an episodic pattern of accretion in the disk. These parameters correspond to the basic model from Article~I, which analyzes it in detail. Over time, matter accumulates in the inner region of the disk, after which this region becomes convectively unstable. In the convectively unstable region, the total viscosity increases by about two orders of magnitude, leading to a relatively rapid discharge of matter onto the star. The middle panel in Fig.~\ref{evolmdot} (left column) shows density distributions for several times, which illustrate this process. The accumulation phase lasts $\sim$3000~years; the convective phase lasts $\sim$250~years.

A further increase in the gas inflow rate from the envelope onto the disk to $10^{-5}$~M$_{\odot}$/yr (Model~3; Fig.~\ref{evolmdot}, bottom panel) leads to an increase in the frequency of accretion bursts and in the maximum accretion rate.
With an increased inflow of matter from the envelope, the accumulation time of matter before the onset of convective instability decreases ($\sim$700~years), while the maximum accretion rate onto the star during the convective phase increases by two orders of magnitude, compared with the basic model. It should also be noted that the minimum accretion rates (between the bursts) in Models 2 and 3 are comparable.

\begin{figure*}
\begin{center}
\hfill
\includegraphics[width=1\columnwidth]{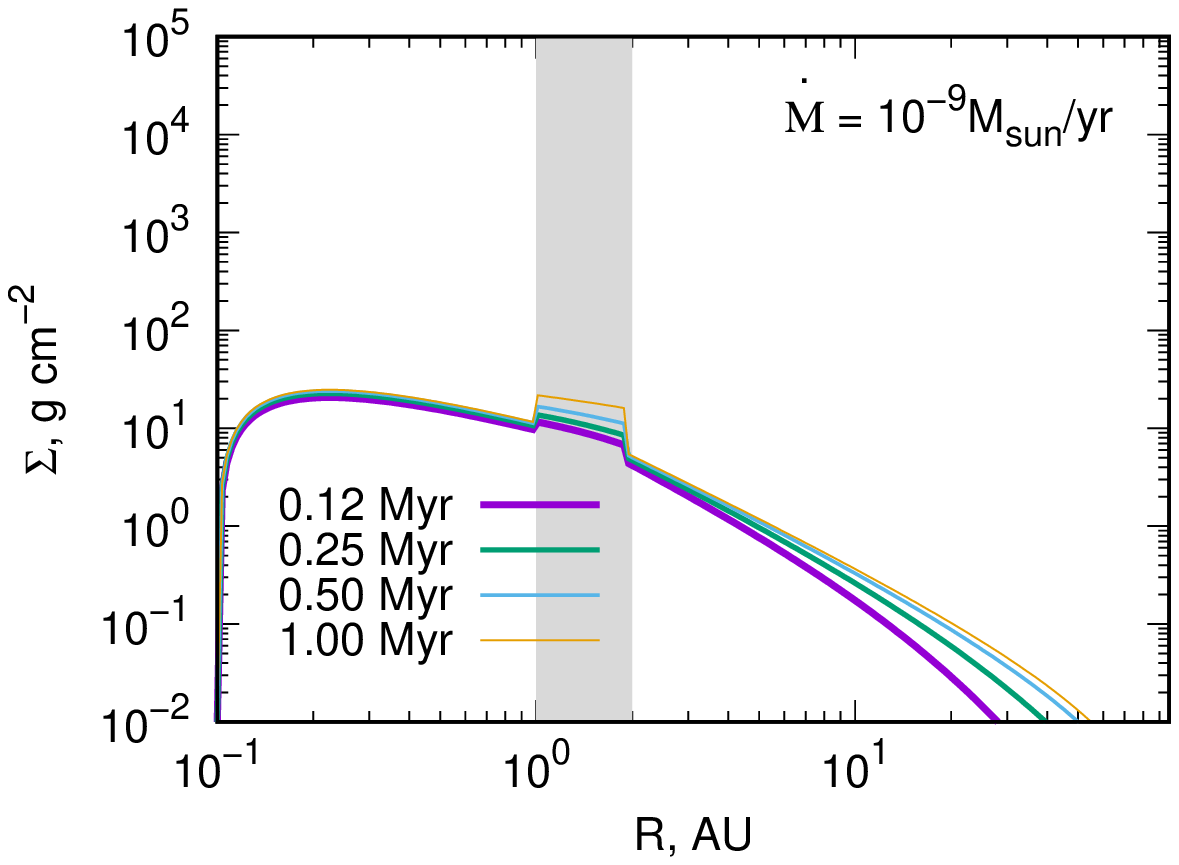}
\includegraphics[width=1\columnwidth]{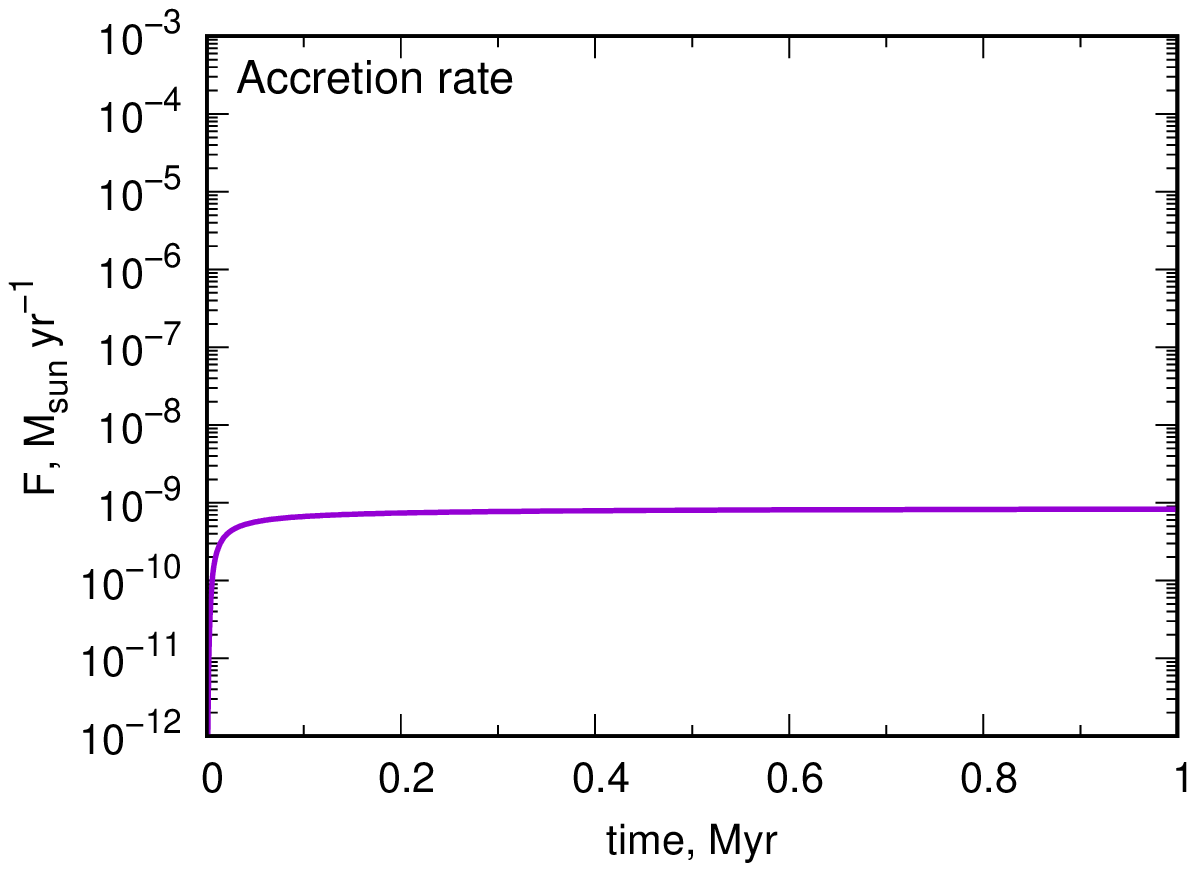}\\
\hfill
\includegraphics[width=1\columnwidth]{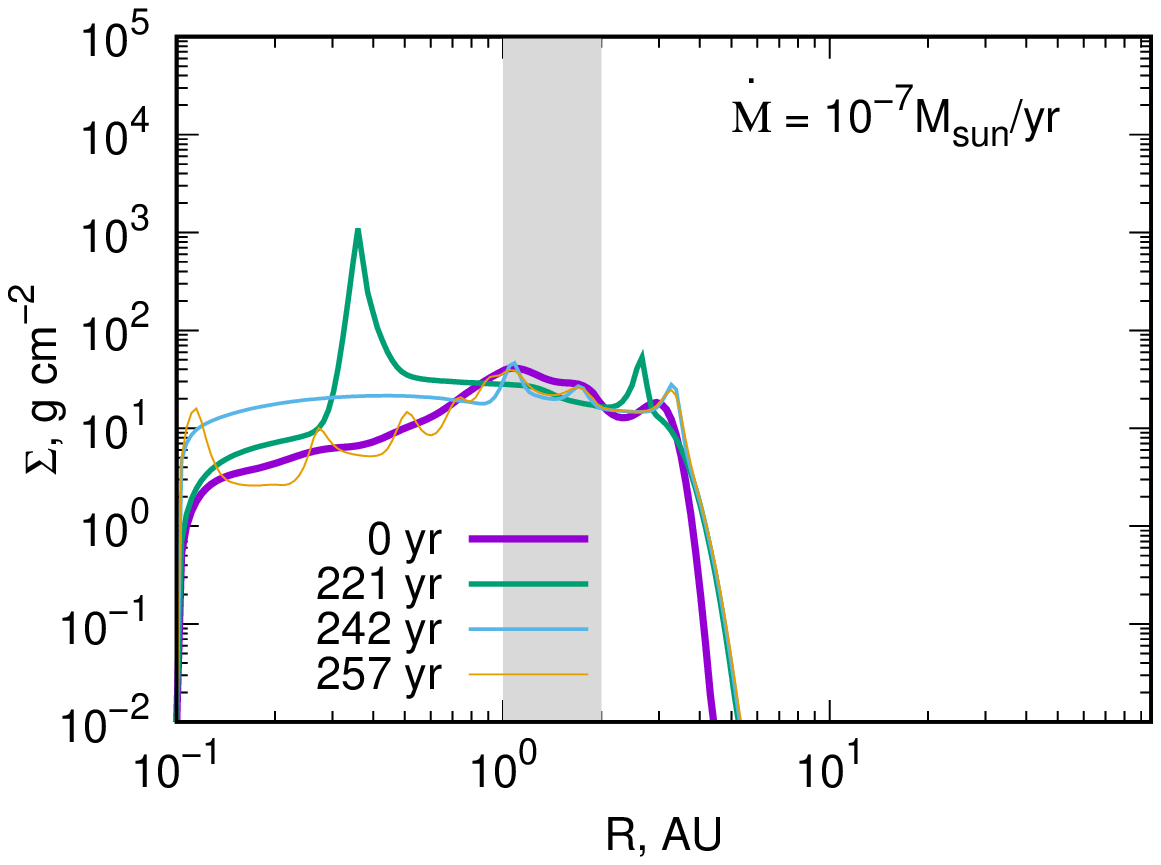}
\includegraphics[width=1\columnwidth]{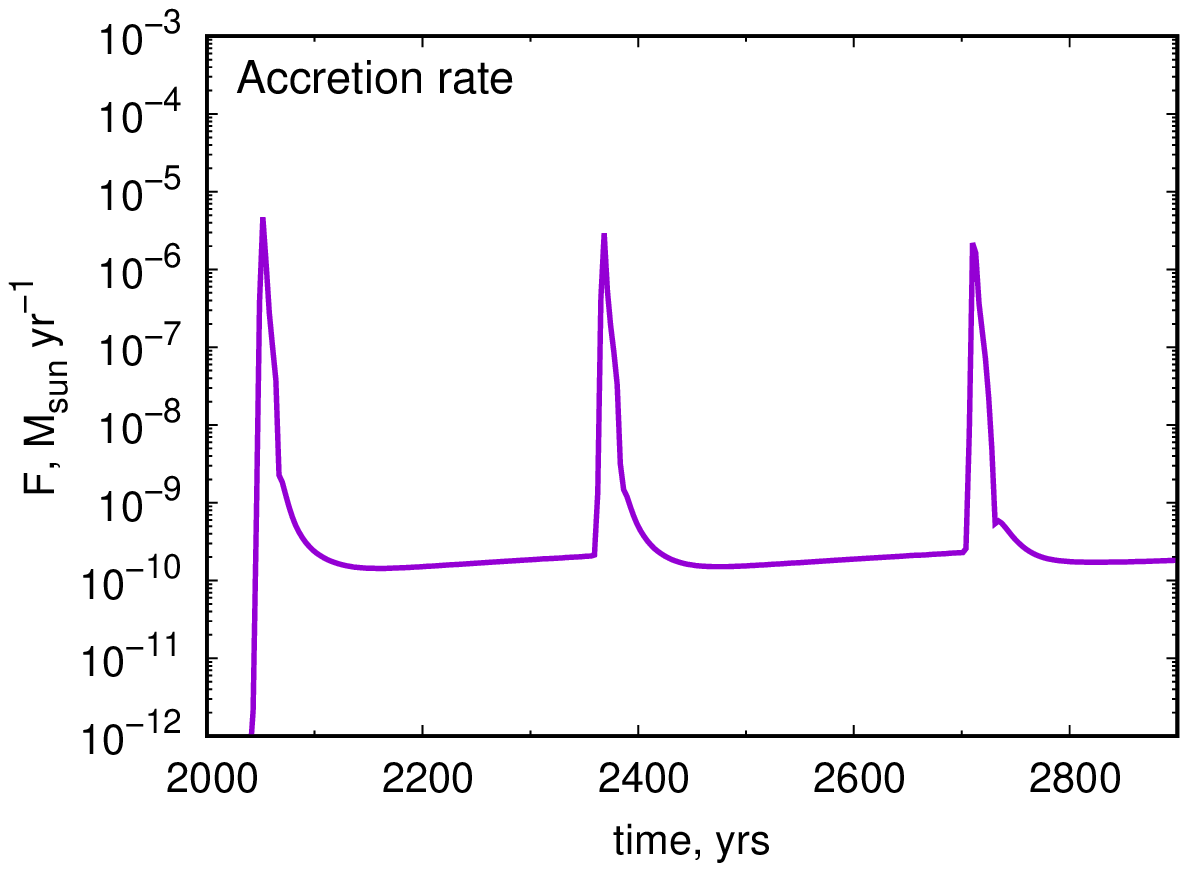}\\
\hfill
\includegraphics[width=1\columnwidth]{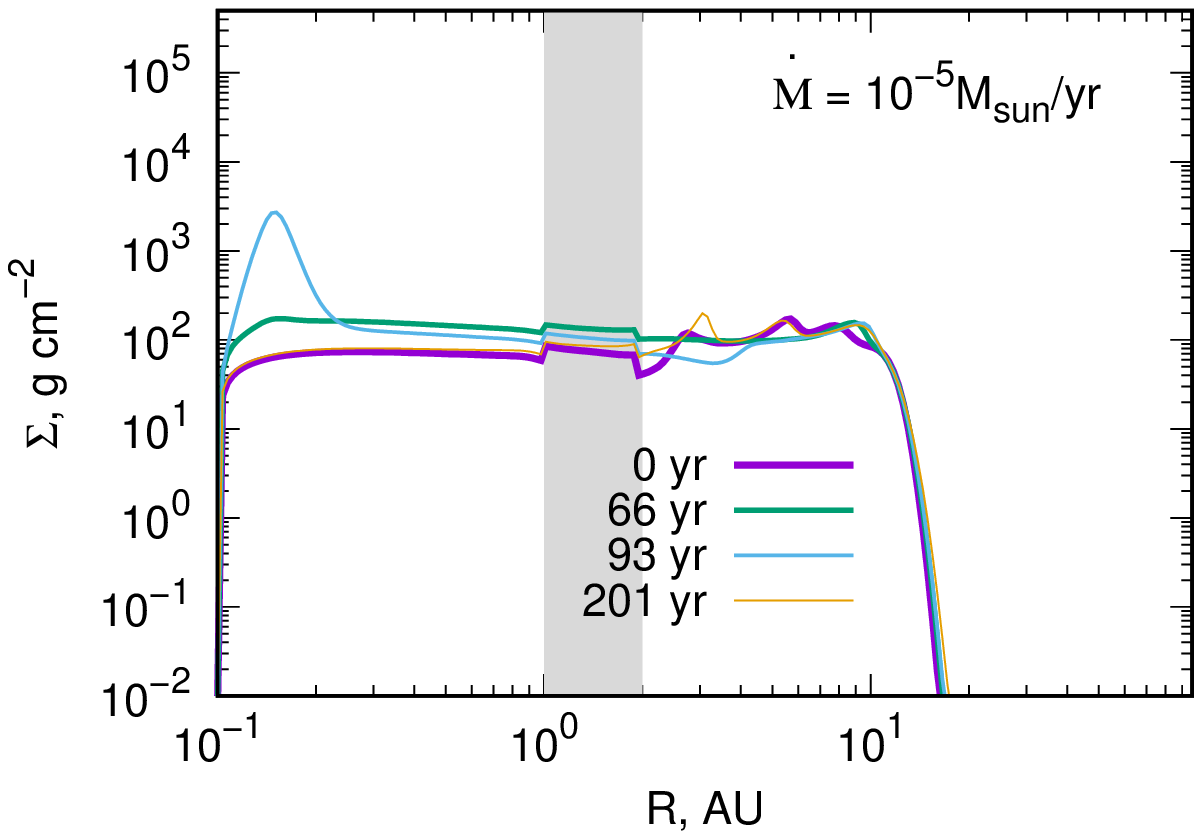}
\includegraphics[width=1\columnwidth]{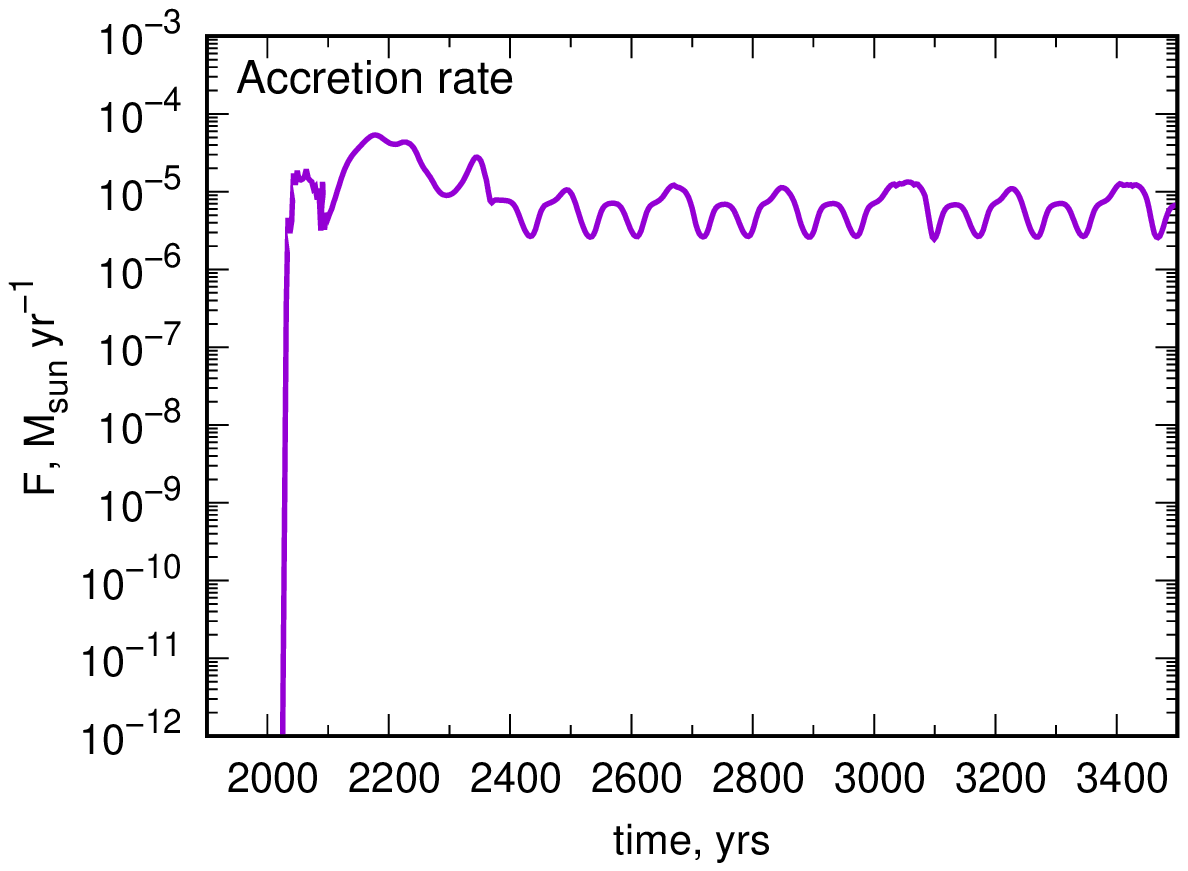}\\
\vspace{1cm}
\caption{The same as in Fig.~\ref{evolmdot} for Models~4, 5, and 6 with an inner accumulation region (the top, middle, and bottom panels, respectively).}
\label{evolmdot2au}
\end{center}
\end{figure*}

\subsection{Models with an Inner Accumulation Region}
Here, we consider the results for the models with
an inner accumulation region ($R_\text{ring}=1-2$~AU, 
$\nu_{0}=10^{15}$~cm$^2$/s, $\eta=1$) and different rates of accretion from the envelope: $\dot{M}=10^{-9}$, $10^{-7}$, and $10^{-5} M_{\odot}$/yr (Models~4, 5,
and 6, respectively). The radial distributions of surface
density and the rates of matter accretion from the disk
onto the star for this model set are shown in Fig.~\ref{evolmdot2au}. In
the case of $\dot{M}=10^{-9} M_{\odot}$/yr (Model~4), it is evident
that convectively unstable regions do not appear. In
complete analogy with Model~1, the density distributions have features near the disk boundaries and the accumulation region.

Let us highlight several differences between
Models~5 and 2, which both have an external inflow
with a rate of $10^{-7} M_{\odot}$/yr. First, we should note a
decrease in the accumulation phase duration by about
an order of magnitude (to $\sim$250~years) and a reduced
convective phase (the burst lasts approximately one
fifth of the time in the basic model, i.e., 50~years
instead of 250). Second, in Model~5, the maximum
level of matter accretion from the disk onto the central
object is an order of magnitude lower. We should also
note that the burst-like pattern of matter accretion
onto the star sets in faster in the case of an inner accumulation
region.

There are far more differences between Models~3
and 6, with $\dot{M}=10^{-5} M_{\odot}$/yr. In Model~6, the inner
region has not enough time to release the accumulated
matter and constantly remains in a state of convective
instability. Nevertheless, the time dependence of the
accretion rate shows relatively weak oscillations in the
range from $10^{-5}$ to $10^{-6} M_{\odot}$/yr (Fig.~\ref{evolmdot2au}, bottom panel).
These oscillations are due to the unstable outer
boundary of the convective zone, which lies beyond
the accumulation region; in this outer zone, matter accumulates and discharges in the same way as the inner
regions in the basic model of the disk.

\begin{figure*}
\hfill
\includegraphics[width=1\columnwidth]{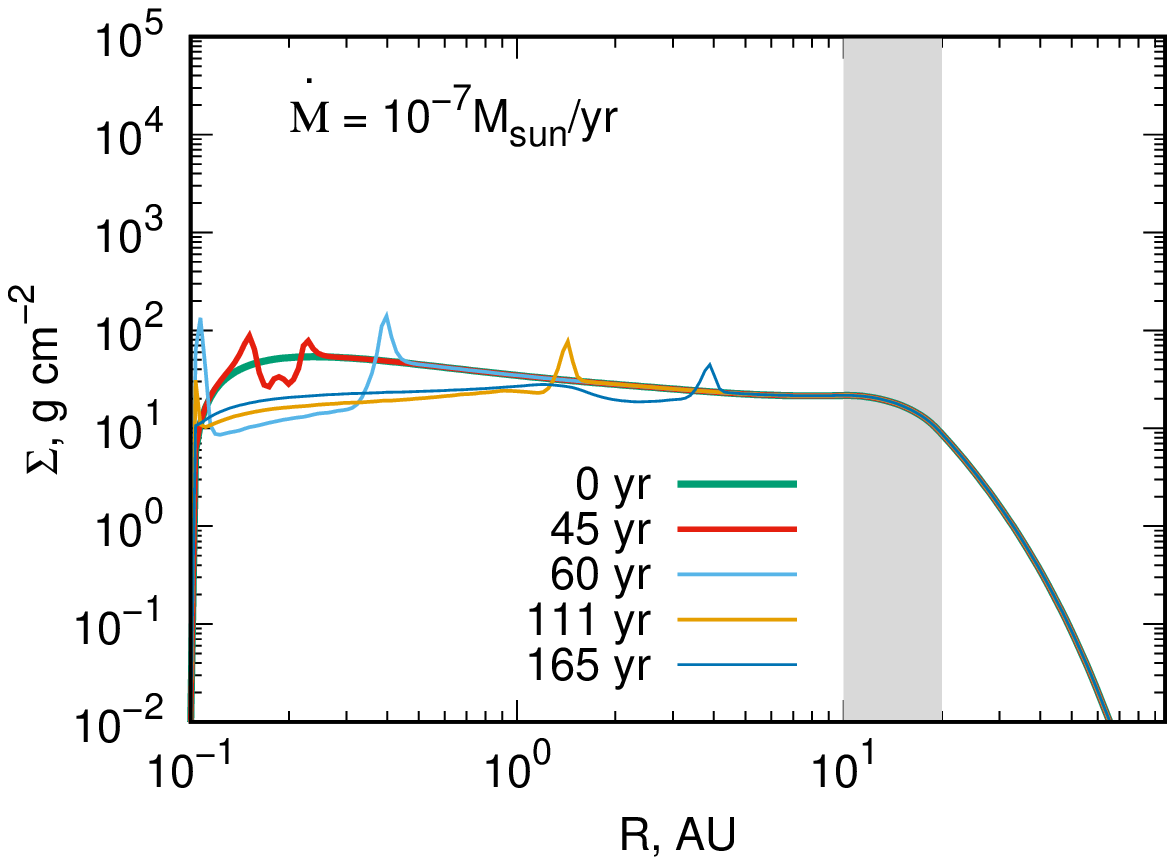}
\includegraphics[width=1\columnwidth]{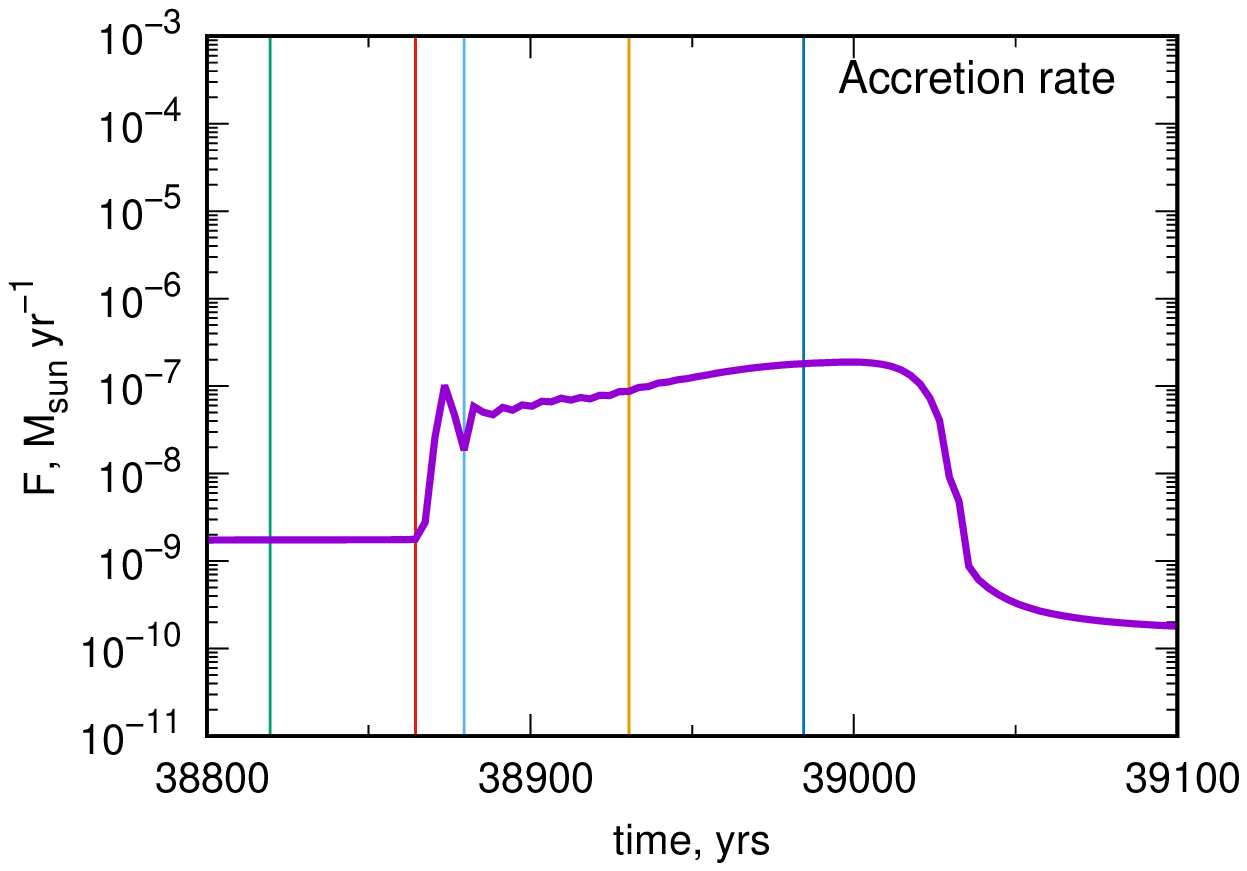}\\
\hfill
\includegraphics[width=1\columnwidth]{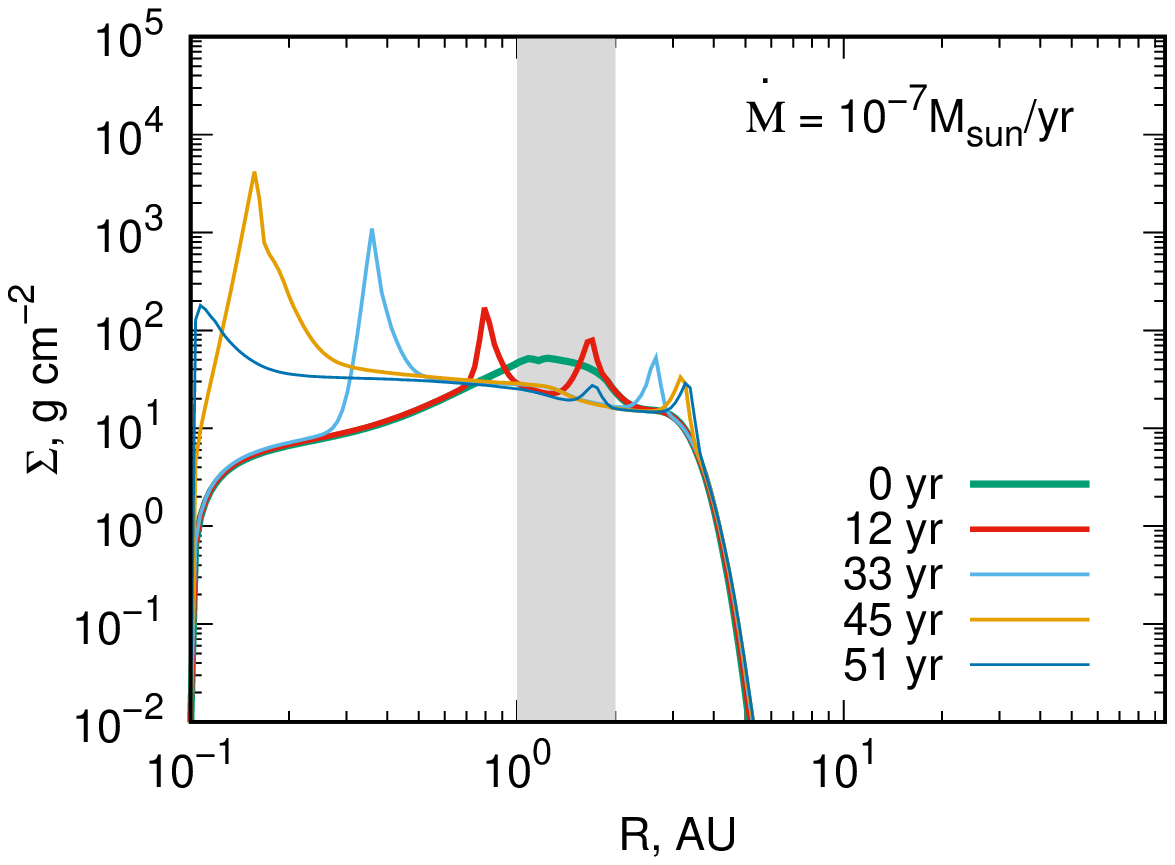}
\includegraphics[width=1\columnwidth]{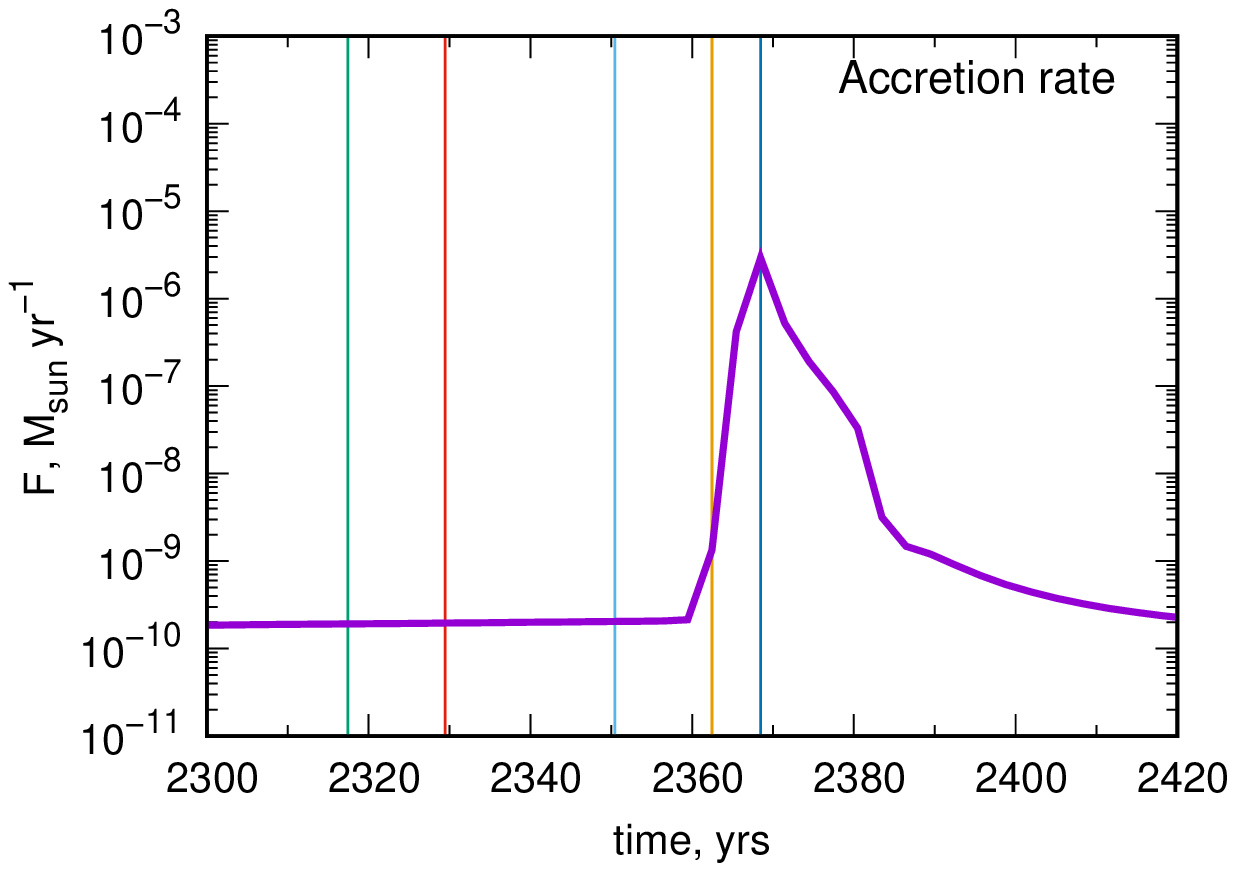}
\vspace{1cm}
\caption{Development of an accretion burst in Model~2 (top panel) and Model~5 (bottom panel). Left: the radial surfacedensity distributions (time is counted from the end of the previous accretion burst). The vertical band shows the gas accretion region from the envelope. Right: the accretion rate of matter from the disk onto the star. The position and color of the vertical lines corresponds to the distributions in the left panel.}
\label{evol_flash}
\end{figure*}

Let us analyze in more detail the development of
bursts in Models~2 and 5. In Figure~\ref{evol_flash}, we show a more detailed evolution of the surface density distributions and the time dependence of the accretion rate in the burst phase. As noted in Article~I, during the convective phase, the density distributions
have peaks, which are essentially the convection
propagation fronts. Looking at these peaks, one can
easily identify the position of the convectively unstable
region. In Model~2, the convectively unstable region
appears near the diskТs inner boundary (0.2~AU) and
expands further  outwards. This development of
the convective region leads to the formation of a $\Pi$-shaped profile of the accretion rate (the top right
panel). In Model~5, however, convection initializes in
the accumulation region and propagates into the diskТs inner
part. This leads to the accumulation of matter on the
inner front and its sharp discharge onto the star, leading to the formation of a $\Lambda$-shaped profile of the burst.
These features of the accretion rate profile may be
important for interpreting observations in young
bursting objects.

\subsection{Effects of the Background Viscosity and the Convection Efficiency Coefficient}

\begin{figure*}
\hfill
\includegraphics[width=1\columnwidth]{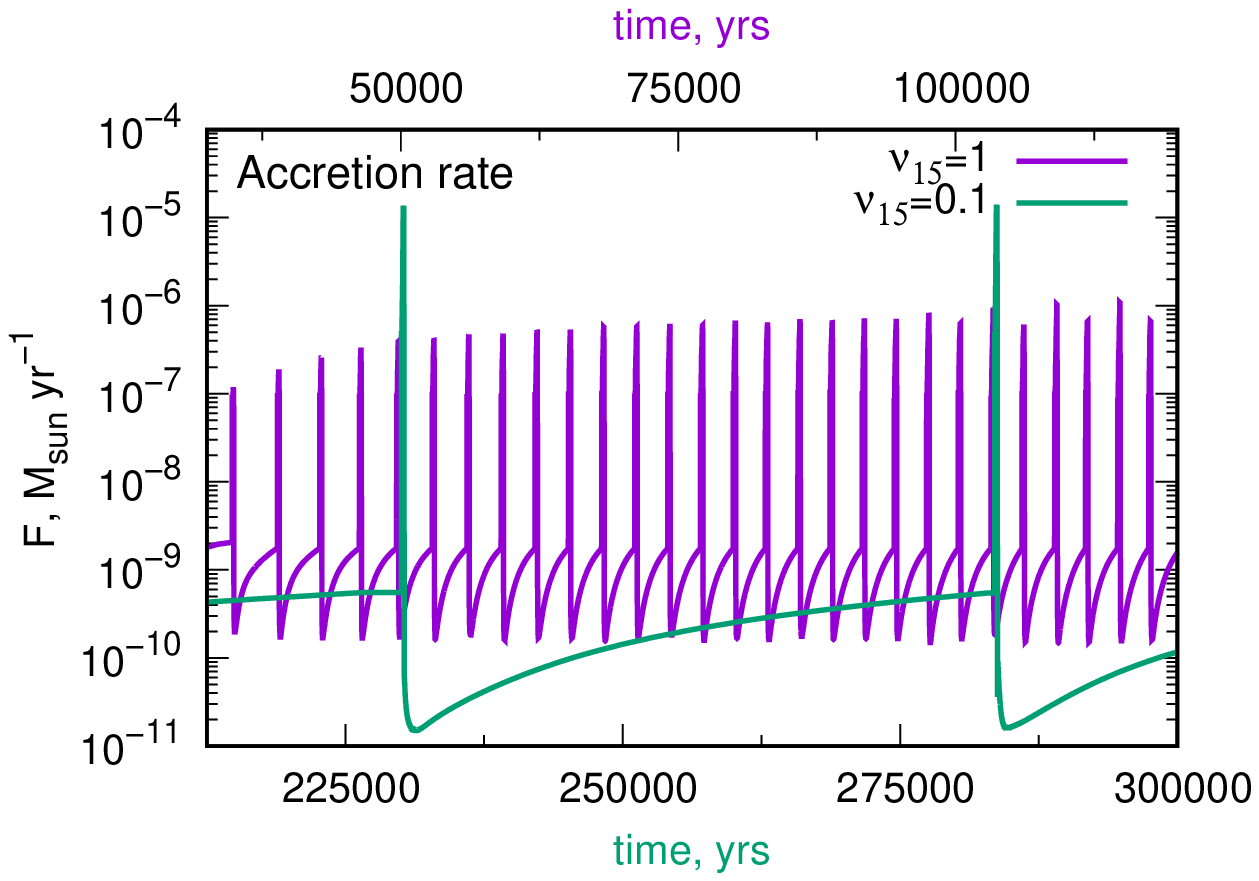}
\includegraphics[height=5.2cm,width=1\columnwidth]{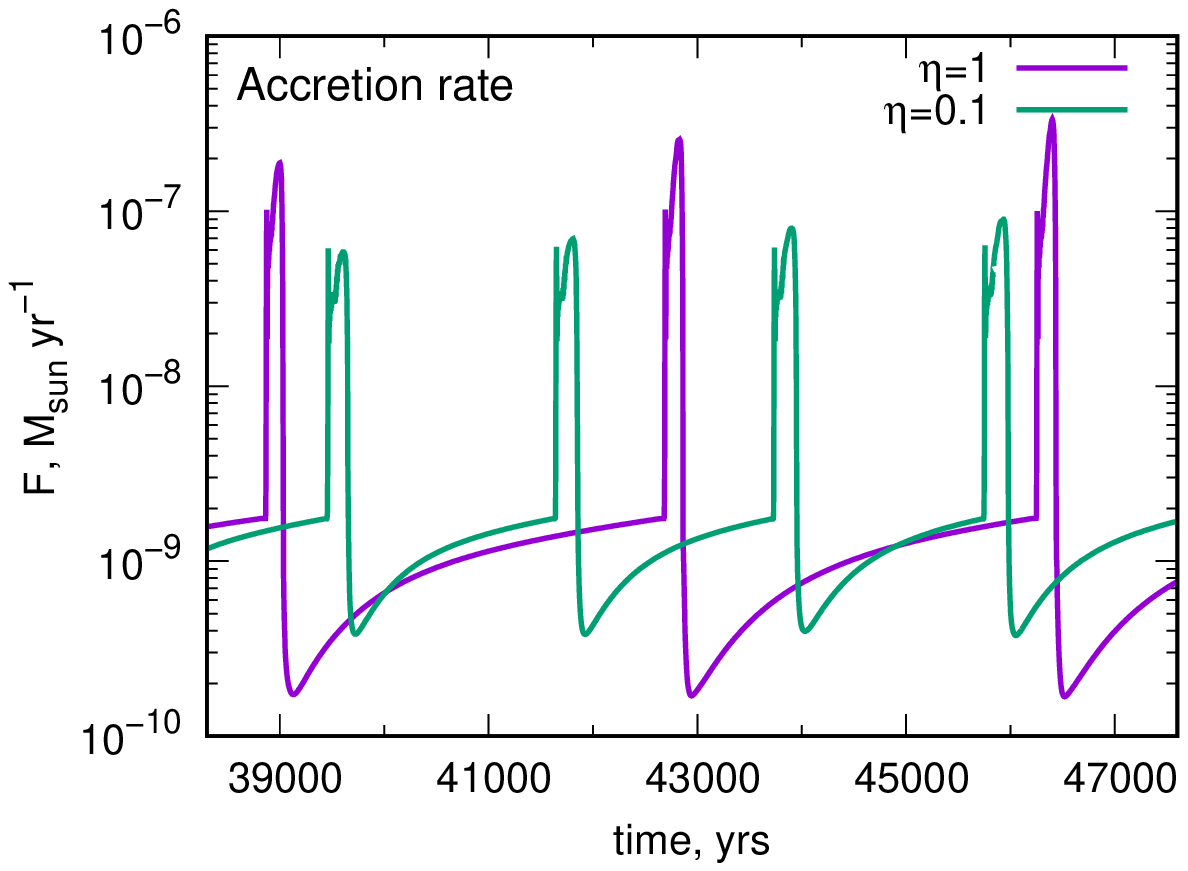}
\vspace{1cm}
\caption{The accretion rate of matter from the disk onto the star for different models. Left: the basic model (Model~2, $\nu_{15}=1$) and
the reduced background viscosity model (Model~7, $\nu_{15}=0.1$). The upper abscissa axis corresponds to the basic model; the lower
one, to Model~7. Right: the basic model (Model~2, $\eta=1$) and the model with the reduced convection efficiency coefficient
(Model~8, $\eta=0.1$).}
\label{evolqeff}
\label{evol_nu}
\end{figure*}

Here, we consider the results of Model~7 (Fig.~\ref{evol_nu},
left panel), in which the background viscosity is an
order of magnitude lower than in the basic model. The
bursts in this model appear at later times (after
225000 yrs) than in the basic one (around 30000 yrs).
In the case of the reduced background viscosity, the
interval between the bursts increases by a factor of
about 20, and so does (i.e., increases by the same factor) the maximum intensity of accretion during the
burst. In Model~7, the minimum accretion rate
decreases tangibly, to $10^{-11} M_{\odot}$/yr. These features are
due to the reduced background viscosity causing a
lower rate of viscous dissipation, which enables the
accumulation of large masses in the disk before the
onset of convective instability. Thus, the decrease in
the background viscosity does not make the bursts disappear but transforms them into less frequent yet more
intensive events. It should be noted that the durations
of the bursts in Models~2 and 7 are comparable.
During the convective phase, the main contribution to
the viscosity coefficient $\nu(R,t)$ comes from the convective viscosity $\nu_{c}$ (see formula~(\ref{eq_nu_all})), which does not depend on $\nu_{0}$.

The right panel of Fig.~\ref{evol_nu} shows the accretion rates
for the basic model and for the one with the reduced convection efficiency coefficient. Evidently, the decrease in $\eta$ does not make the bursts disappear either, but it modifies them. The bursts become a quarter more frequent yet less intensive. The reduced convection efficiency leads to a lower convective viscosity coefficient  $\nu_{\rm c}(R,t)$ (see formulae ~(\ref{eq_nuconv})-(\ref{eq_gammavis})). Since the convective viscosity decreases, the convective phase becomes less intensive. Thus, the disk discharges less mass during the burst, which leads to smaller intervals between the bursts.

Note that the model calculations presented here are for illustrative purposes only. Their main goal is to demonstrate the possible role of convection in achieving a nonstationary accretion regime in accretion disks and to qualitatively assess the importance of certain parameters. In the above described Models~1Ц8, accretion was set constant in time and space. We performed simulations until the onset of bursts if any appeared, as well as studied their characteristics. However, the further evolution of the disk was not investigated. In fact, both the accretion rate and the fall area from the envelope should be changing over time. In the next section, we modeled the long-term evolution of the disk with the view of this dependence.

\section{EVOLUTION OF PDs UNDER A VARIABLE INFLOW OF MATTER FROM THE ENVELOPE}
\label{sec_evol}
In order to investigate the long-term evolution of
the disk, we need to specify a realistic function $W(R,t)$, which describes the rate of matter inflow from
the envelope. To calculate this function, we use an
approximation about the conservation of matterТs
local angular momentum in the accreting envelope,
i.e., the residue of the parent protostellar cloud.
Within this approximation, an element of volume
originally located at a distance of $l$ from the polar axis
falls onto the so-called centrifugal radius $R_c$:
\begin{equation}
R_c=\dfrac{l^4 \Omega^2}{GM_{*}},
\label{Racc}
\end{equation}
at which its angular velocity becomes Keplerian. In
this relation, $\Omega$ is the initial angular velocity of the element under consideration and $M_{*}$ is the current stellar mass. Thus, the model assumes that the cloud elements settle gradually onto the disk at a local Keplerian velocity, with each cloud element having its own
settling radius, which is calculated from the angular
momentumТs conservation condition for the cloud
element.

\begin{figure*}
\hfill
\includegraphics[width=0.66\columnwidth]{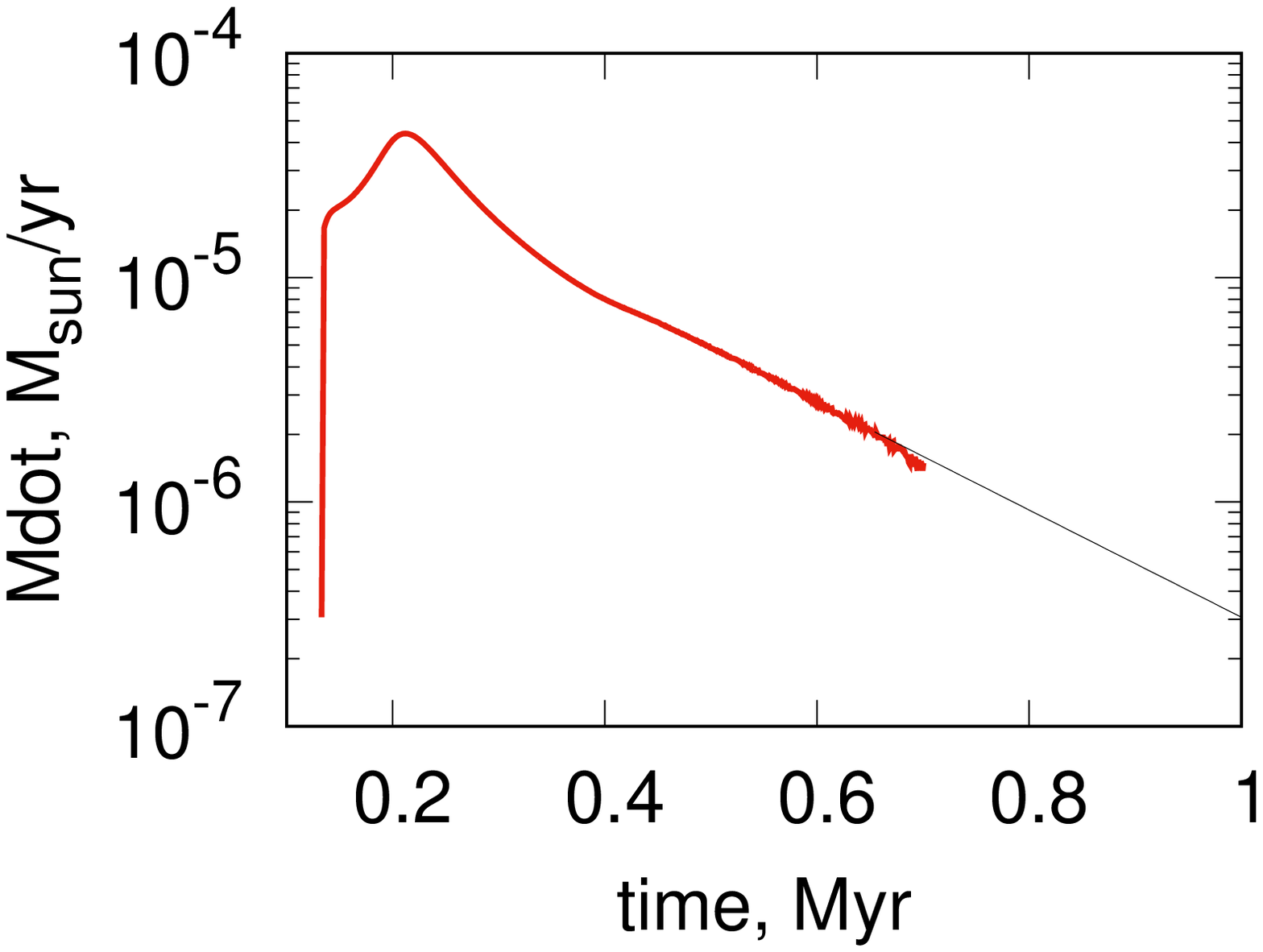}
\includegraphics[width=0.66\columnwidth]{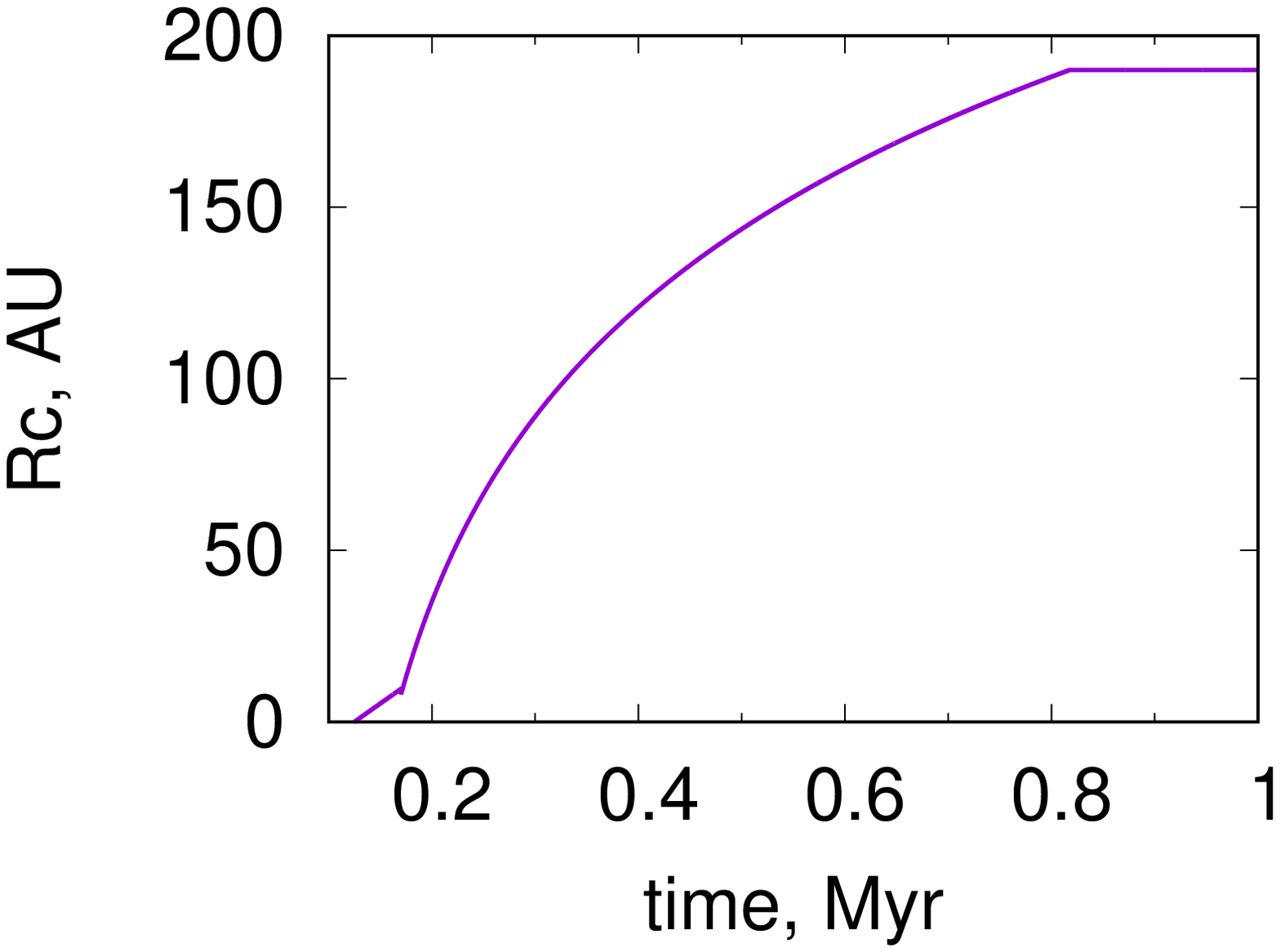}
\includegraphics[width=0.66\columnwidth]{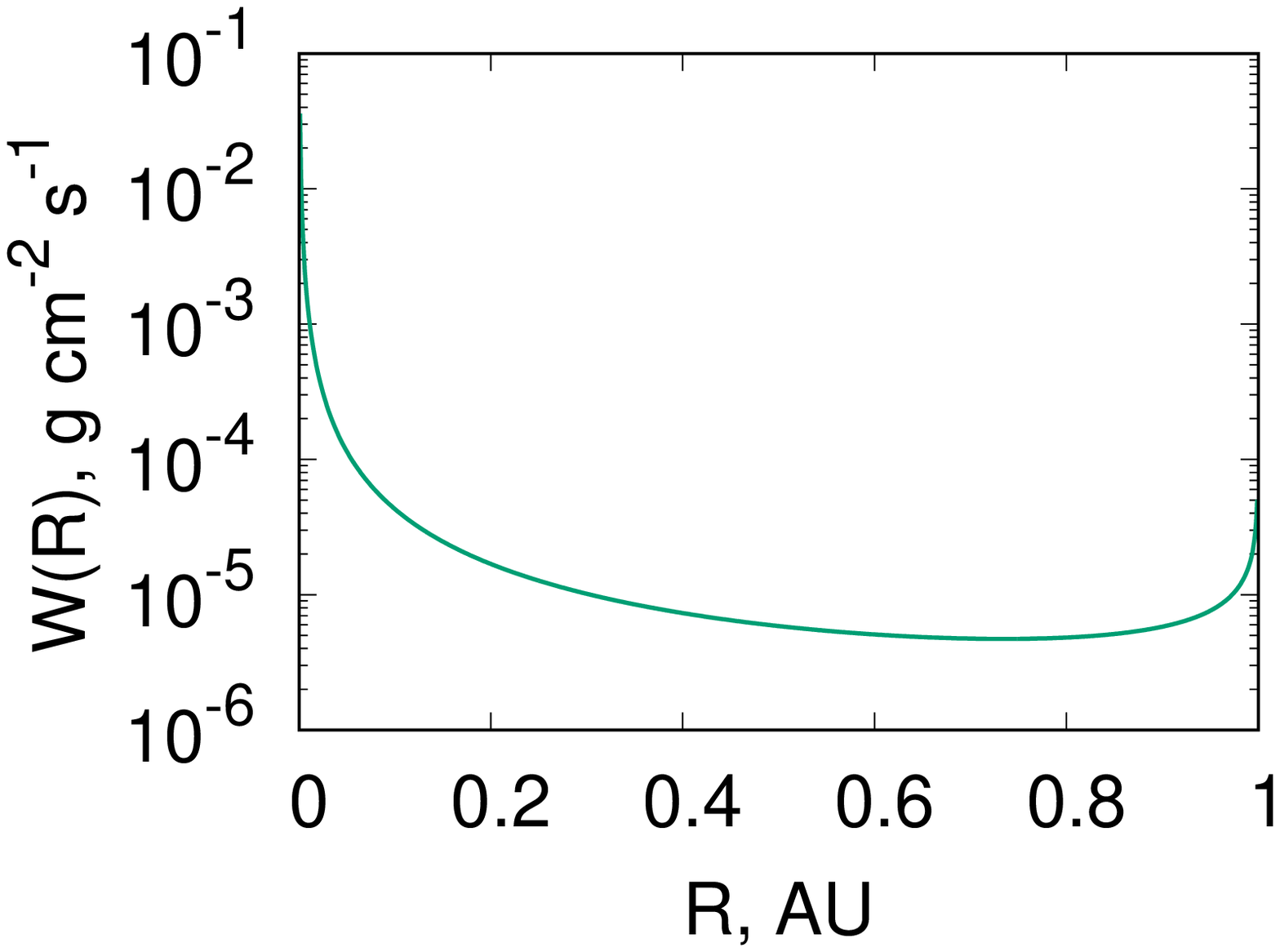}
\vspace{1cm}
\caption{The accretion rate from the envelope onto the disk. The black line shows the extrapolation of the numerical solution (left);
the dependence of the maximum centrifugal radius $R_{c}$ on time (center) and the shape of the $W(R,t)$ function for an accretion rate of $\dot{M}=10^{-4}M_{\odot}$/yr and $R_{c}=1$~AU (right).}
\label{ev_rc}
\end{figure*}

If we assume that the original protostellar cloud is
spherically symmetric and rotates in a solid-body
manner, then the function $W(R,t)$ is written as~\citep{2005A&A...442..703H}
\begin{equation}
W(R,t) = \frac{\dot{M}(t)}{8\pi R_c^{2}(t)}  \left(\frac{R}{R_c(t)}\right)^{-3/2}
\left[1-\left(\frac{R}{R_c(t)}\right)^{1/2}\right]^{-1/2},
\label{W_rt}
\end{equation}
where $\dot{M}(t)$ is the current full accretion rate from the
envelope onto the disk; $R_c(t)$ is the boundary of the
accumulation region, i.e., the centrifugal radius for an
accreted element from the equatorial plane. The functions $\dot{M}(t)$ and $R_c(t)$ can be specified using different
approaches~\citep{2005A&A...442..703H}. We take these functions by approximating and extrapolating the numerical simulation
results of a cloud collapse and subsequent accretion of
the envelope onto the star from~\citep{2015ARep...59..133P}. We can rightfully do so because aforementioned work used the Lagrange method, which creates a framework to trace the evolution of
individual elements. In so doing, we used the
angular velocity of the original cloud $\Omega=10^{-14}$~s$^{-1}$,
which is a characteristic quantity for the cores of
molecular clouds~\citep{2013EAS....62...25B}. Figure~\ref{ev_rc} presents the functions $\dot{M}(t)$ and $R_c(t)$ and shows the shape of the function $W(R,t)$.

The accretion rate from the envelope onto the disk
in the range 0.2--0.7~Myr is well approximated by an
exponential function. As we lack data on the subsequent evolution of the envelope, we use this approximation for larger times as well. The centrifugal radius
increases with time, reaching $\approx$180~AU at 0.8~Myr. At
large times, we use a constant quantity of 180~AU. It is
evident from the shape of the function $W(R,t)$ that the
matter falling onto the disk fills unevenly in an area
inside the accumulation region. Specifically, the maximum of
$W(R,t)$ near zero is due to the fall of matter onto the
disk from the envelopeТs circumpolar regions.

Within this model, we study the diskТs long-term
evolution; therefore, we need to consider that the stellar mass increases due to the matter inflow from the
disk. We do so by assuming that the initial stellar mass
is 0.3~$M_{\odot}$ and increase it in accordance with the
accreted mass. The radius and photospheric luminosity of the star should also change simultaneously with
its mass, but we neglect this change for the sake of simplicity, assuming that the stellar radius and luminosity
are equal to those of the Sun. As in the constant inflow
model, the accretion luminosity of the central object is
variable and is calculated from~\eqref{acc_lum}. Note that it is the
accretion luminosity that makes the greatest contribution to the luminosity of the central object at early stages in the disk evolution. The coefficients $\nu_{0}=10^{15}$~cm$^2$/s and $\eta=1$ were taken from the basic model and were time independent.

\begin{figure*}
\begin{center}
\includegraphics[width=0.99\columnwidth]{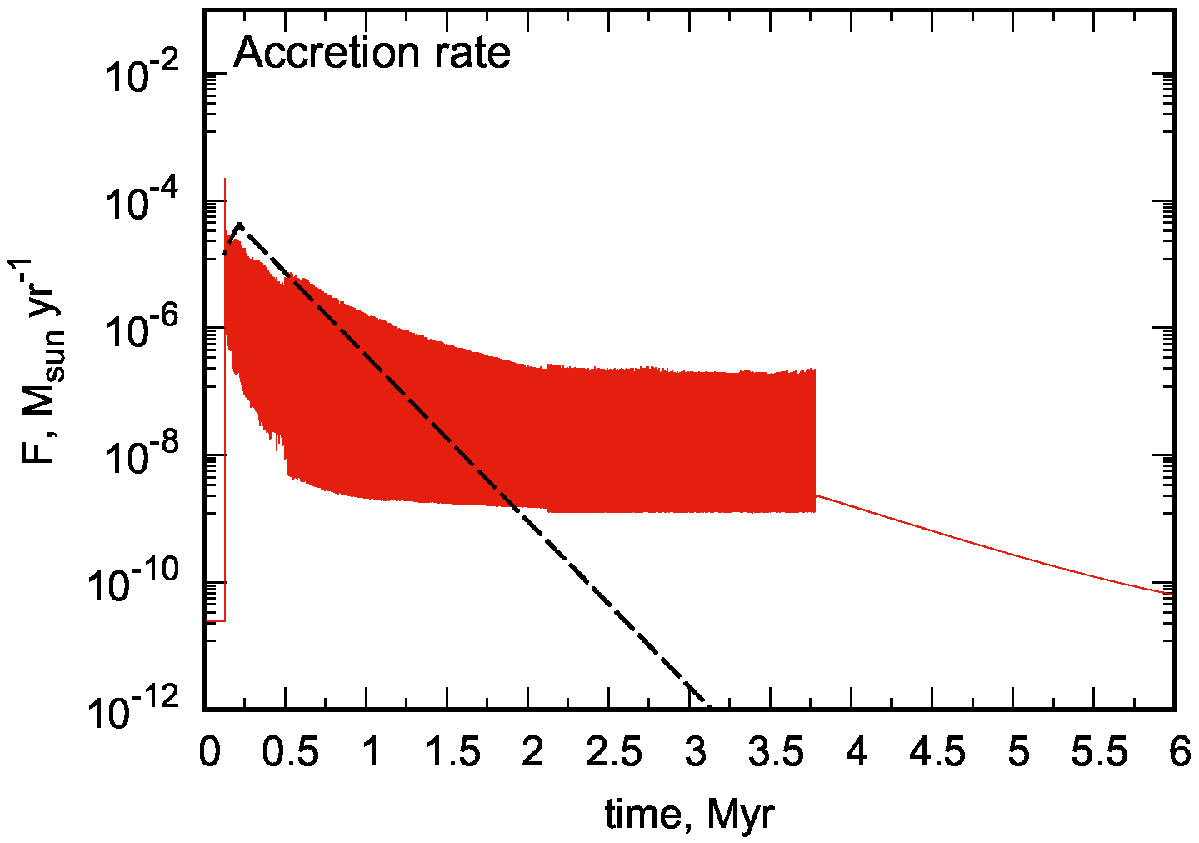}
\hspace{0.2cm}
\includegraphics[width=0.99\columnwidth]{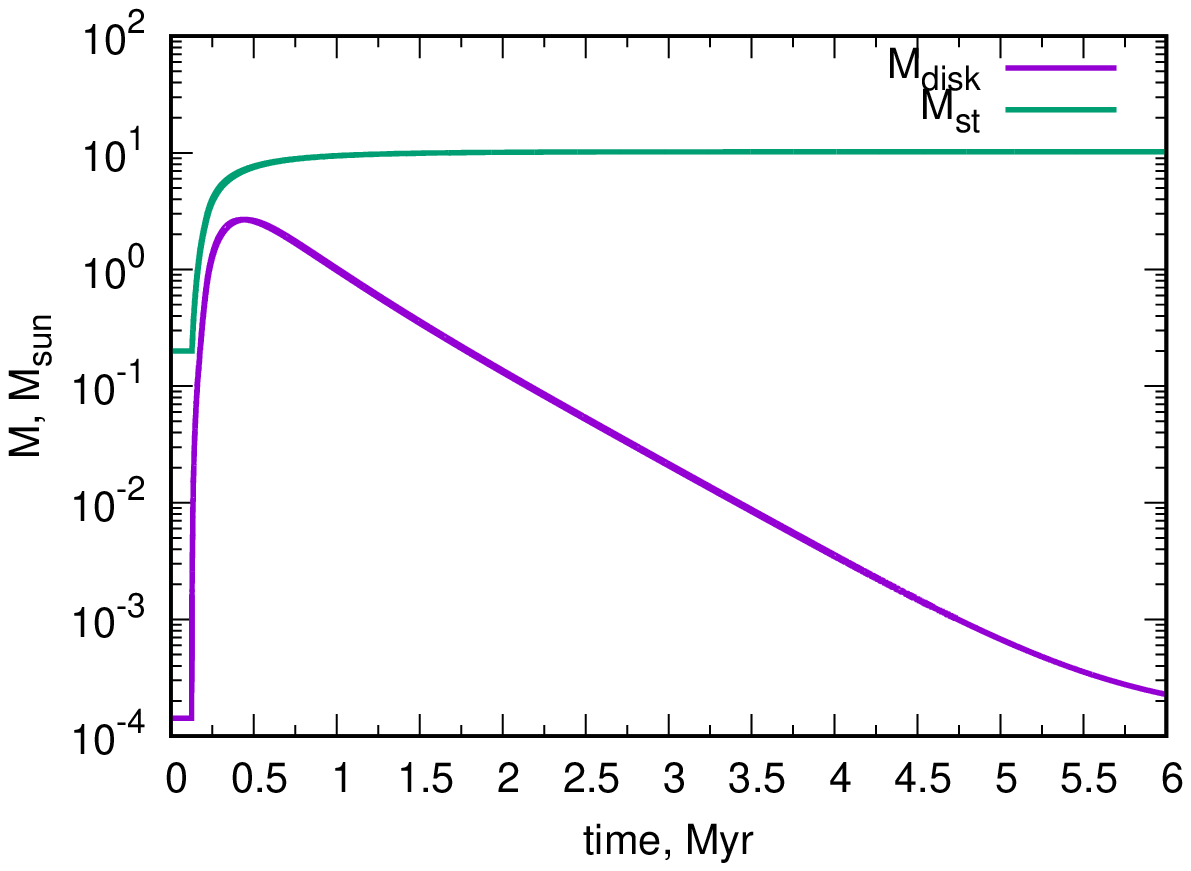}
\caption{Top: Accretion rate evolution of matter from the disk onto the star (the dashed black line indicates a given matter inflow
from the envelope onto the disk). Bottom: the change in the disk mass and stellar mass with time.}
\label{ev_accr}
\end{center}
\end{figure*}

Let us consider the calculated results for the disk
evolution in this model. Figure~\ref{ev_accr} shows the time
dependence of the accretion rate from the disk onto
the star and the change in the disk mass with time. The
filled area in the accretion rate distribution at 0.17--3.7~Myr indicates a burst-like accretion regimeЧat
this figure scale, the numerous bursts merge into a single continuous band. After 3.7~Myr, the bursts cease to
occur, and the accretion rate smoothly decreases with
time. A comparison between the accretion rate onto
the star and the rate of matter inflow from the envelope (the dashed line in Fig.~\ref{ev_accr}, top panel) leads to a
conclusion about the importance of the disk mass
accumulation process. In the first million years, the
disk accumulates a considerable mass (see Fig.~\ref{ev_accr}, bottom
panel), and its subsequent evolution is defined by the
redistribution of this mass while the matter inflow
from the envelope becomes negligibly small. In the
burst phase, both the maximum and minimum accretion rates decrease smoothly over $\approx 2$~Myr, after which
they remain virtually constant until 3.7~Myr. We
should also note that the change in the accretion rate
in the quiet phase ($t>3.7$~Myr) agrees well with the
analytical dependence $\dot M \propto t^{-5/4}$, which describes a
disk with a viscosity distribution of $\nu \propto R$ (see formula (6) from~\citep{2004ARep...48..800T}).

\begin{figure*}
\includegraphics[width=0.66\columnwidth]{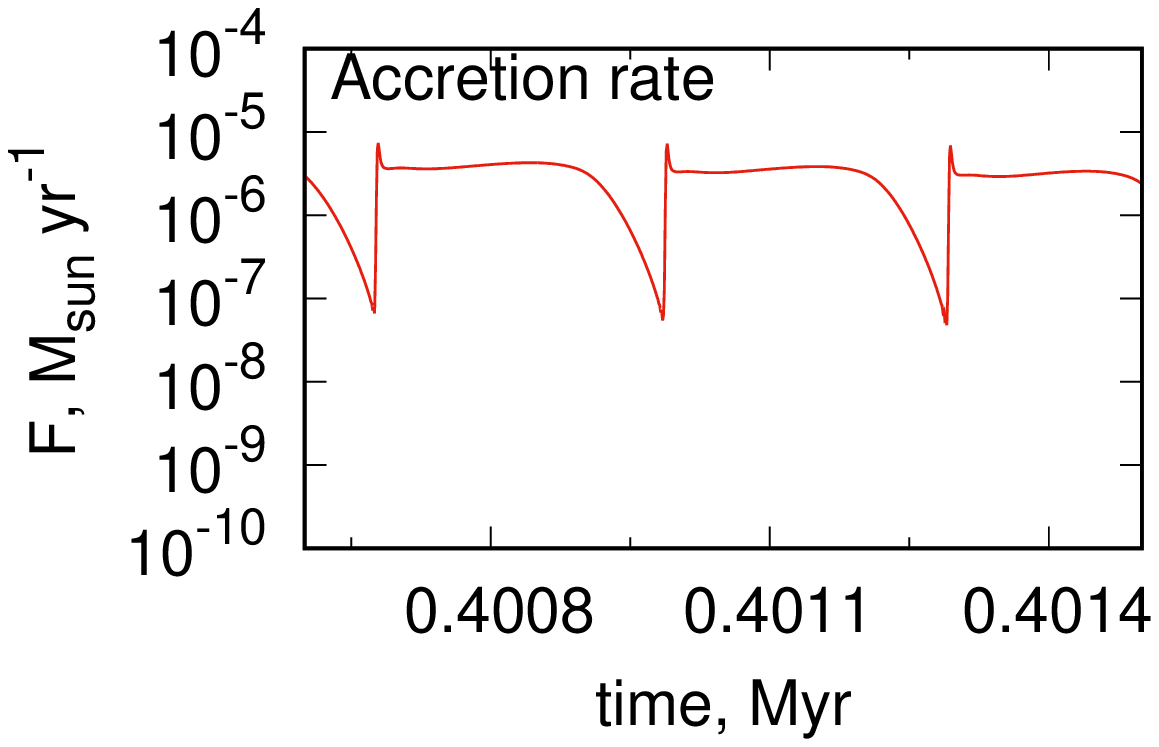}
\includegraphics[width=0.66\columnwidth]{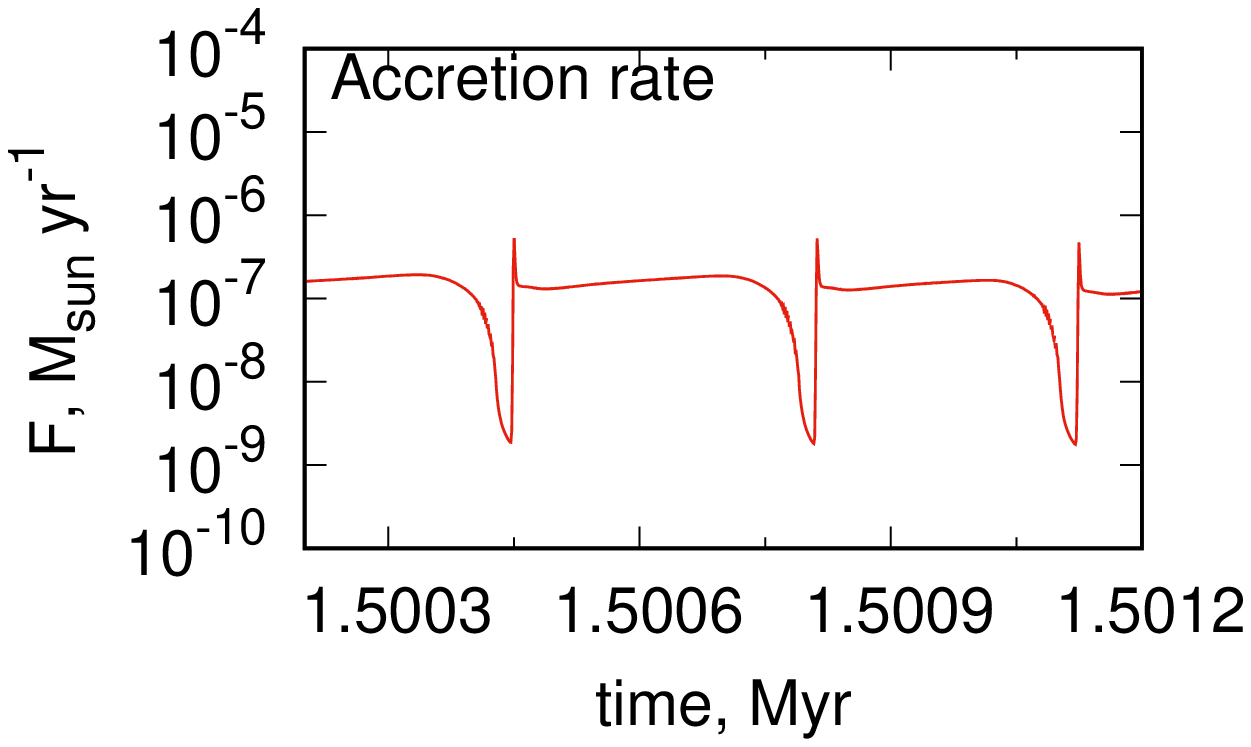}\includegraphics[width=0.66\columnwidth]{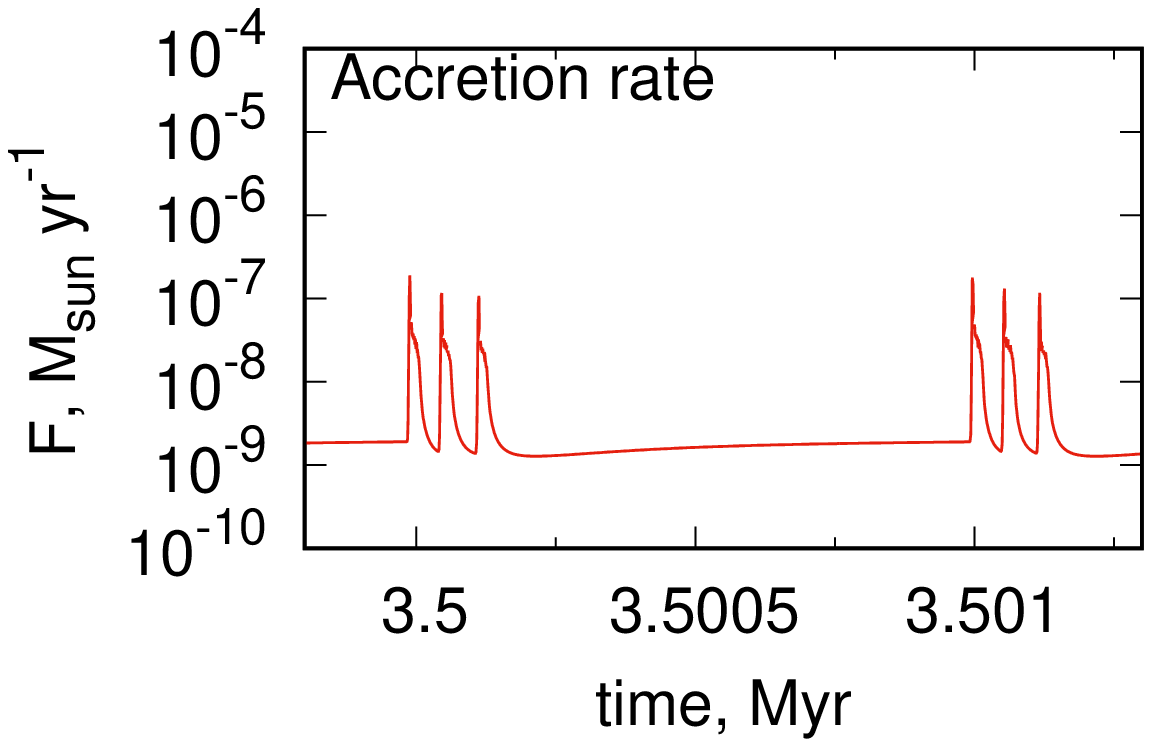}
\vspace{1cm}
\caption{The accretion rate of matter from the disk onto the star for three time intervals in the neighborhood of 0.4~Myr (left),
1.5~Myr (center), and 3.5~Myr (right).}
\label{flashes}
\end{figure*}

Figure~\ref{flashes} shows characteristic forms of accretion
bursts at times in the neighborhood of 0.4, 1.5, and
3.5~Myr. Evidently, the bursts have profiles differing
from those described by us for the constant inflow
model. Specifically, the bursts at 0.4 and 1.5~Myr have
deep and narrow minima directly before the maximum. Meanwhile, the bursts at 3.5~Myr are morphologically similar to those described in Section~\ref{sec_param} but are
of composite nature. These differences are due to the
advanced evolution of the disk and the influence of its
outer parts, i.e., the mass reservoir for convectively
unstable regions, which was left out of consideration in
the constant inflow model.

\begin{figure}
\includegraphics[width=0.99\columnwidth]{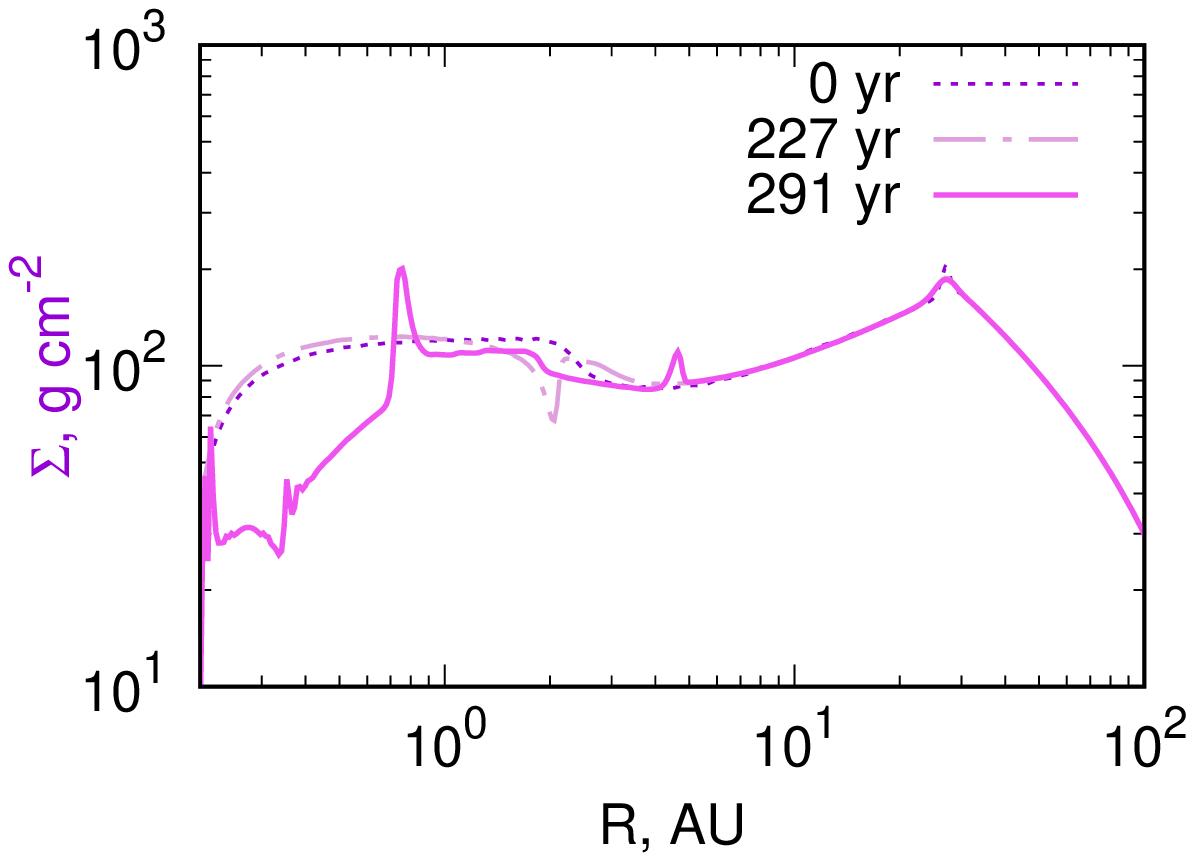}\\
\includegraphics[width=0.99\columnwidth]{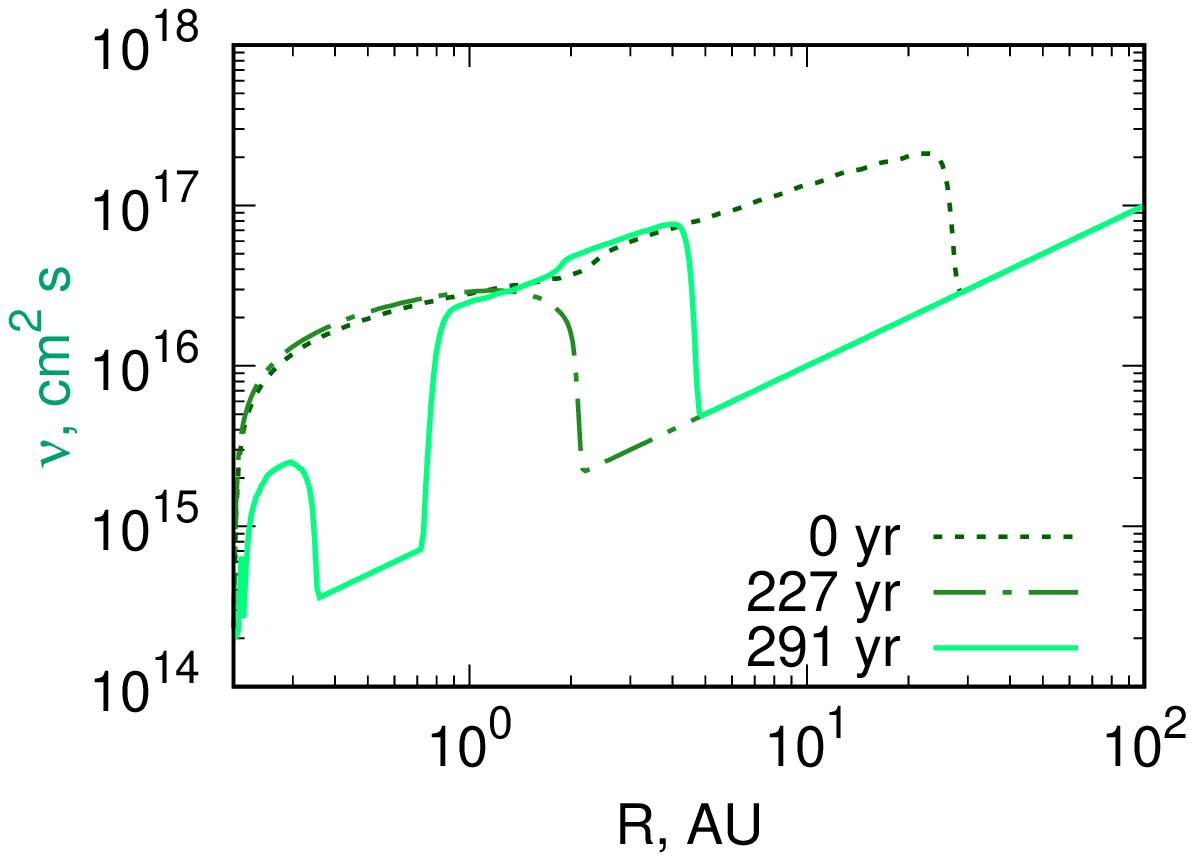}
\vspace{1cm}
\caption{Top: the radial surface-density distribution for three times in the neighborhood of 1.5~Myr. Bottom: the total viscosity coefficient $\nu(R,t)$ for the same times.}
\label{4_nuW}
\end{figure}

Let us analyze the formation of a burst at a time
interval in the neighborhood of 1.5~Myr as an example.
Figure~\ref{4_nuW} presents the diskТs surface density and the
total viscosity coefficient $\nu(R)$ for three neighboring
times. At the conventionally initial time, the entire
inner region up to 30~AU is convective, which is evident from the high viscosity coefficient (Fig.~\ref{4_nuW}, right
panel). Over time, the size of the convective zone
decreases; i.e., its boundary shifts towards the star,
reaching 2~AU at time 227~yrs. At time 291~yrs, the
outer boundary of this convective zone contracts to a
radius of 0.35~AU and will soon reach the diskТs inner
boundary. It is evident that by this time, a new convective zone has formed inside 0.7--5~AU. This new convective zone expands in both directions and will subsequently capture the entire inner zone up to 30~AU.
Thus, the new convective phase in the disk begins to
develop before the previous one comes to an end. It is
the short space interval between the convective zone
boundaries (the interval between 0.35 and 0.7~AU in
Fig.~\ref{4_nuW}) that creates the deep narrow minimum before
the accretion maximum.

\begin{figure}
\includegraphics[width=0.99\columnwidth]{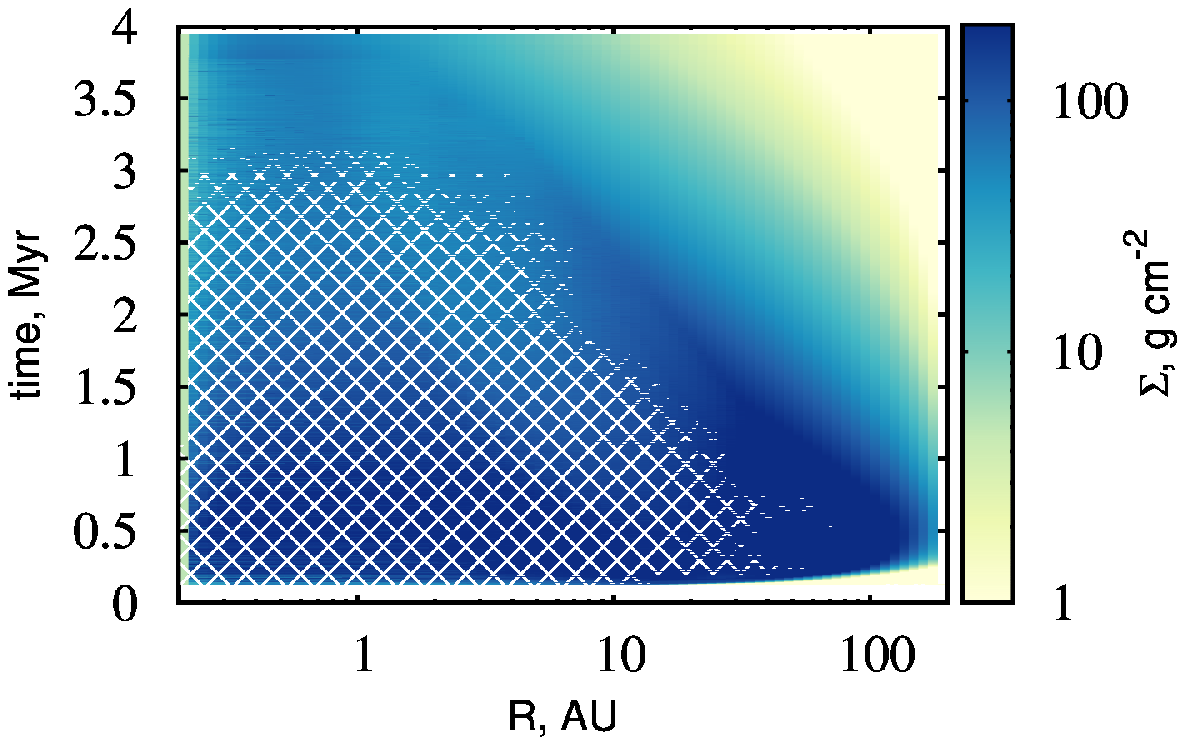}\\
\includegraphics[width=0.99\columnwidth]{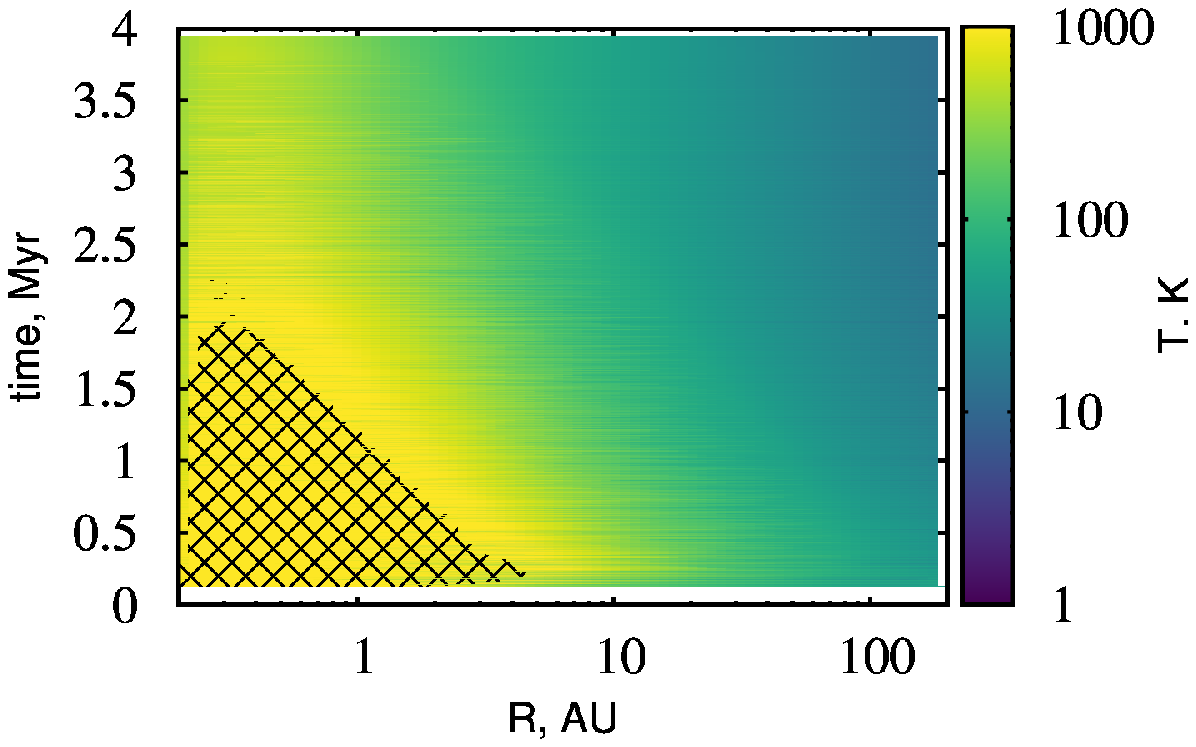}\\
\includegraphics[width=0.99\columnwidth]{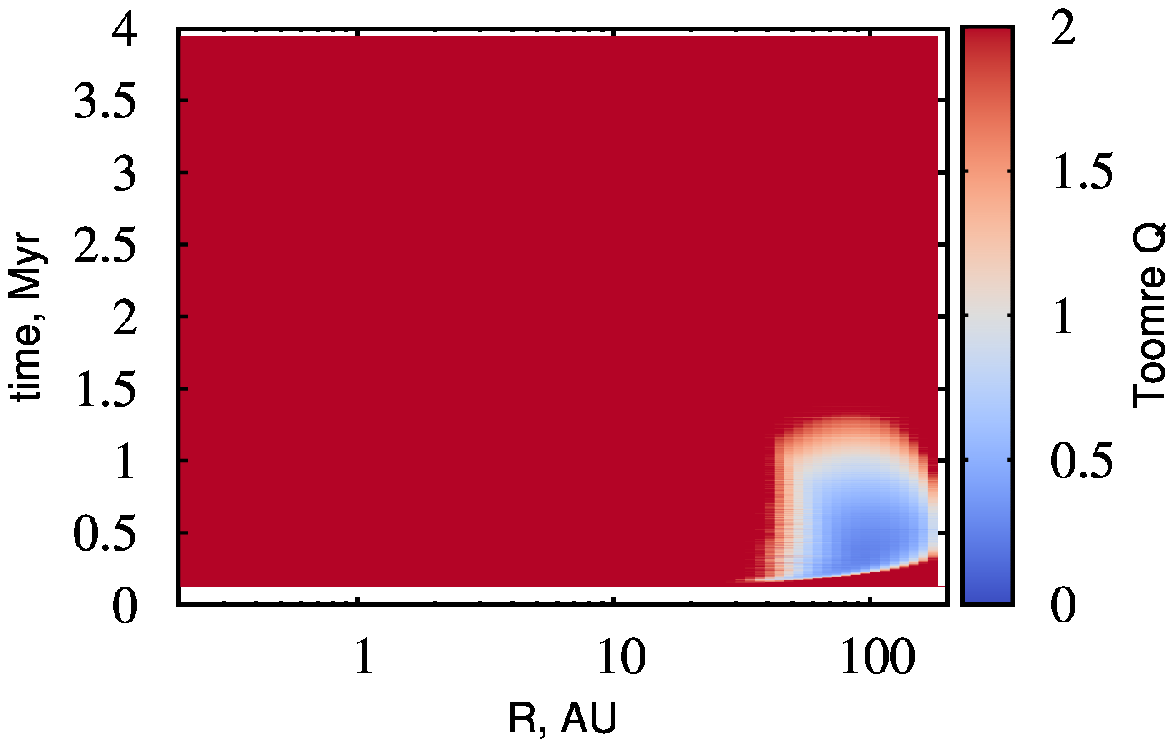}
\caption{Top panel: the long-term evolution of the radial surface-density distribution. The crosshatching indicates the
regions that became convectively unstable. Middle panel:
the evolution of the radial equatorial-temperature distribution. The crosshatching indicates the regions where
temperature exceeded 1500~K at its maximum. Bottom
panel: the distribution of the Toomre parameter $Q$. The
values of $Q>2$ are shown in red. The values of $Q<1$
(shades of blue) indicate gravitationally unstable regions.
}
\label{ev_map}
\end{figure}

The above results suggest that convection can serve
as an important factor underlying the nonsteady pattern of accretion from the disk onto the star. Figure~\ref{ev_map} (top panel) shows the long-term evolution of the surface densityТs radial distribution and marks convectively unstable regions. Evidently, in the disk evolution
process, the size of the convectively unstable region
decreases from several tens to several astronomical
units, while the phase of episodic accretion lasts less
than 4~Myr. These results are of qualitative nature;
however, our model has several serious limitations,
which are noted in Article~I. Removing these limitations may largely complicate the disk evolution pattern. One of these limitations is that the model
neglects the dust evaporation and gas dissociation/ionization processes occurring at high temperatures. Figure~\ref{ev_map} (middle panel) shows the evolution of
the equatorial temperature distribution and marks the
regions with temperatures above 1500~K, in which the
dust evaporation processes become important. Obviously, these regions concentrate in the diskТs inner
parts and manifest themselves more conspicuously at
initial times. They are seen to partially overlap the
convectively unstable regions, which situation should
of course affect the disk evolution pattern. Meanwhile, the convectively unstable region is wider in
space and time, which is why the conclusions from the
model under consideration remain relevant.

Another limitation of the model resides with its
neglect of the diskТs self-gravitation (Eq.~\eqref{eq1} is valid for
the Keplerian disk). However, as seen from Fig.~\ref{ev_accr}, the
disk mass in the early stages of evolution is comparable
with the stellar mass. Figure~\ref{ev_map} (bottom panel) shows
the distribution of the Toomre parameter~\citep{1960AnAp...23..979S,1964ApJ...139.1217T}: $Q=\dfrac{c_s\Omega}{\pi G \Sigma}$, where $c_s$ is the velocity of sound and $\Omega$ is the Keplerian angular velocity. Low values of this parameter ($Q<1$) indicate gravitationally unstable regions. Evidently, these regions appear at initial times of the evolution ($t<0.5$~Myr) in the diskТs outer parts ($R>50$~AU). The appearance of these regions should also
affect the disk evolution; i.e., the disk should develop
spirals and fragments, whose interaction with one
another and with the disk creates complex variations
(see, e.g.,~\citep{2017A&A...606A...5V}). Thus, at initial times in the evolution, convection may be complicated by other, possibly
more complex, governing processes.

\section{CONCLUSIONS}
\label{sec_final}

According to Article~I, convective instability can
lead to nonsteady accretion onto the star in a PD.
However, this conclusion was there illustrated only by one
model with a fixed set of parameters. In this work, we
investigated how the model parameters affect the
onset of episodic accretion and studied the pattern of
this accretion at different matter inflow rates from the
envelope and for different accumulation regions in the disk.
The results from these models can be summarized as
follows:
\begin{itemize}
\item Depending on the matter inflow rate from the
envelope onto the disk, we can identify three main
accretion regimes: (a) low inflow: no convection
occurs and accretion is monotonic; (b) moderate inflow:
convectively unstable regions arise periodically, which
leads to nonsteady (burst-like) accretion; and (c) high
inflow: the diskТs inner regions may become completely convective, which leads to a weakly oscillating
pattern of accretion due to instabilities behind the
accumulation region of the disk.

\item Burst parameters (maximum intensity, duration,
and period) depend on the matter inflow rate and the
position of the accumulation region. Thus, with an increase in
the external inflow, we see an increase in the intensity
and frequency of the bursts. A shift of the accumulation region
towards the center leads to an increased frequency and
decreased duration of the bursts.

\item The onset of episodic accretion is a stable manifestation of the disk model used in the study. Specifically, if the background viscosity and convection efficiency are reduced by an order of magnitude, this
change does not make the bursts disappear yet modifies them.
\end{itemize}
In addition to this analysis, we modeled the longterm evolution of the disk, including a variable matter
inflow from the envelope $W(R,t)$. This calculation
enabled us to trace the disk evolution from the first
luminosity bursts to their complete termination and
gradual depletion of the disk. Based on the calculation
results, we made the following conclusions:
\begin{itemize}
\item An important effect of the disk evolution from
the perspective of periodic accretion manifestations is
the mass accumulation process in the disk due to matter inflow from the envelope. In the first million years,
the disk accumulates considerable mass, and the subsequent disk evolution is defined by the redistribution
of this mass rather than by accretion from the envelope, which becomes negligible.

\item The disk soon becomes convectively unstable
and remains so for almost 4~Myr. Meanwhile, the
instability captures an area of several tens of astronomical units and then gradually decreases.

\item Burst parameters (intensity, duration, and frequency), as well as their shape, change with time,
which is associated with a change in the disk mass and the integral flow of matter through it. The bursts may
take very bizarre shapes.
\end{itemize}
We also illustrated the model limitations Ч the calculations provide conditions for a gravitational instability to arise as well as for high-temperature regions
where dust evaporation is expected. These processes
are neglected in the model. Moreover, the model has
several other limitations, which were detailed in Article~I. Therefore, the results presented are largely qualitative in nature. Specifically, in the early phases of the
disk evolution, convection can coexist with other, possibly more intense, processes. We believe that further
investigation of the convection role should rely on
more consistent models, which will consider hydrodynamic effects, dust evaporation, and gas dissociation
and ionization processes.

\section*{Acknowledgments}

The authors thank the reviewer for valuable comments
and constructive suggestions for improving this article.

The work by L.A. Maksimova was carried out in the framework of the project "Study of stars with exoplanets" under a grant from the Government of the Russian Federation for scientific research conducted under the guidance of leading scientists (agreement N 075-15-2019-1875).

\bibliographystyle{mn2e} 
\bibliography{paper}

\begin{thebibliography}{30}
\expandafter\ifx\csname natexlab\endcsname\relax\def\natexlab#1{#1}\fi

\bibitem[{{Audard} {et~al}\mbox{.}(2014){Audard}, {{\'A}brah{\'a}m}, {Dunham},
  {Green}, {Grosso}, {Hamaguchi}, {Kastner}, {K{\'o}sp{\'a}l}, {Lodato},
  {Romanova}, {Skinner}, {Vorobyov}, \& {Zhu}}]{2014prpl.conf..387A}
{Audard} M. {et~al.}, 2014, in Protostars and Planets VI, {Beuther} H.,
  {Klessen} R.~S., {Dullemond} C.~P., {Henning} T., eds., p. 387

\bibitem[{{Balbus} \& {Hawley}(1991)}]{1991ApJ...376..214B}
{Balbus} S.~A., {Hawley} J.~F., 1991, \apj, 376, 214

\bibitem[{{Belloche}(2013)}]{2013EAS....62...25B}
{Belloche} A., 2013, in EAS Publications Series, Vol.~62, EAS Publications
  Series, {Hennebelle} P., {Charbonnel} C., eds., pp. 25--66

\bibitem[{{Coleman} {et~al}\mbox{.}(2016){Coleman}, {Kotko}, {Blaes}, {Lasota},
  \& {Hirose}}]{2016MNRAS.462.3710C}
{Coleman} M.~S.~B., {Kotko} I., {Blaes} O., {Lasota} J.~P., {Hirose} S., 2016,
  Monthly Notices Roy. Astron. Soc., 462, 3710

\bibitem[{{Ercolano} {et~al}\mbox{.}(2014){Ercolano}, {Mayr}, {Owen},
  {Rosotti}, \& {Manara}}]{2014MNRAS.439..256E}
{Ercolano} B., {Mayr} D., {Owen} J.~E., {Rosotti} G., {Manara} C.~F., 2014,
  Monthly Notices Roy. Astron. Soc., 439, 256

\bibitem[{{Hartmann}(1998)}]{1998apsf.book.....H}
{Hartmann} L., 1998, {Accretion Processes in Star Formation}

\bibitem[{{Hartmann} \& {Kenyon}(1996)}]{1996ARA&A..34..207H}
{Hartmann} L., {Kenyon} S.~J., 1996, Annual Review of Astronomy and
  Astrophysics, 34, 207

\bibitem[{{Hawley} \& {Balbus}(1991)}]{1991ApJ...376..223H}
{Hawley} J.~F., {Balbus} S.~A., 1991, \apj, 376, 223

\bibitem[{{Held} \& {Latter}(2018)}]{2018MNRAS.480.4797H}
{Held} L.~E., {Latter} H.~N., 2018, Monthly Notices Roy. Astron. Soc., 480,
  4797

\bibitem[{{Hirose} {et~al}\mbox{.}(2014){Hirose}, {Blaes}, {Krolik}, {Coleman},
  \& {Sano}}]{2014ApJ...787....1H}
{Hirose} S., {Blaes} O., {Krolik} J.~H., {Coleman} M. S.~B., {Sano} T., 2014,
  \apj, 787, 1

\bibitem[{{Hueso} \& {Guillot}(2005)}]{2005A&A...442..703H}
{Hueso} R., {Guillot} T., 2005, Astron. and Astrophys., 442, 703

\bibitem[{{Kley} \& {Lin}(1999)}]{1999ApJ...518..833K}
{Kley} W., {Lin} D.~N.~C., 1999, \apj, 518, 833

\bibitem[{{Lin} \& {Papaloizou}(1980)}]{1980MNRAS.191...37L}
{Lin} D.~N.~C., {Papaloizou} J., 1980, Monthly Notices Roy. Astron. Soc., 191,
  37

\bibitem[{{Lipunova} \& {Shakura}(2003)}]{2003AN}
{Lipunova} G.~V., {Shakura} N.~I., 2003, Izvestia Rossiiskoi Academii Nauk,
  Seria fizicheskaya, 67, 322

\bibitem[{{Malanchev}, {Postnov} \& {Shakura}(2017){Malanchev}, {Postnov}, \&
  {Shakura}}]{2017MNRAS.464..410M}
{Malanchev} K.~L., {Postnov} K.~A., {Shakura} N.~I., 2017, Monthly Notices Roy.
  Astron. Soc., 464, 410

\bibitem[{{Natta}, {Testi} \& {Randich}(2006){Natta}, {Testi}, \&
  {Randich}}]{2006A&A...452..245N}
{Natta} A., {Testi} L., {Randich} S., 2006, Astron. and Astrophys., 452, 245

\bibitem[{{Pavlyuchenkov} {et~al}\mbox{.}(2020){Pavlyuchenkov}, {Tutukov},
  {Maksimova}, \& {Vorobyov}}]{2020ARep...64....1P}
{Pavlyuchenkov} Y.~N., {Tutukov} A.~V., {Maksimova} L.~A., {Vorobyov} E.~I.,
  2020, Astronomy Reports, 64, 1

\bibitem[{{Pavlyuchenkov} {et~al}\mbox{.}(2015){Pavlyuchenkov}, {Zhilkin},
  {Vorobyov}, \& {Fateeva}}]{2015ARep...59..133P}
{Pavlyuchenkov} Y.~N., {Zhilkin} A.~G., {Vorobyov} E.~I., {Fateeva} A.~M.,
  2015, Astronomy Reports, 59, 133

\bibitem[{{Pringle}(1981)}]{1981ARA&A..19..137P}
{Pringle} J.~E., 1981, Annual Review of Astronomy and Astrophysics, 19, 137

\bibitem[{{Safronov}(1960)}]{1960AnAp...23..979S}
{Safronov} V.~S., 1960, Annales d'Astrophysique, 23, 979

\bibitem[{{Shakura} \& {Postnov}(2015)}]{2015MNRAS.451.3995S}
{Shakura} N., {Postnov} K., 2015, Monthly Notices Roy. Astron. Soc., 451, 3995

\bibitem[{{Shakura}(1972)}]{1972AZh....49..921S}
{Shakura} N.~I., 1972, Astronomicheskii Zhurnal, 49, 921

\bibitem[{{Shakura} \& {Sunyaev}(1973)}]{1973A&A....24..337S}
{Shakura} N.~I., {Sunyaev} R.~A., 1973, Astron. and Astrophys., 24, 337

\bibitem[{{Toomre}(1964)}]{1964ApJ...139.1217T}
{Toomre} A., 1964, \apj, 139, 1217

\bibitem[{{Tutukov} \& {Pavlyuchenkov}(2004)}]{2004ARep...48..800T}
{Tutukov} A.~V., {Pavlyuchenkov} Y.~N., 2004, Astronomy Reports, 48, 800

\bibitem[{{Vorobyov} \& {Basu}(2006)}]{2006ApJ...650..956V}
{Vorobyov} E.~I., {Basu} S., 2006, \apj, 650, 956

\bibitem[{{Vorobyov} \& {Pavlyuchenkov}(2017)}]{2017A&A...606A...5V}
{Vorobyov} E.~I., {Pavlyuchenkov} Y.~N., 2017, Astron. and Astrophys., 606, A5

\bibitem[{{Vorobyov} {et~al}\mbox{.}(2019){Vorobyov}, {Skliarevskii},
  {Elbakyan}, {Pavlyuchenkov}, {Akimkin}, \& {Guedel}}]{2019A&A...627A.154V}
{Vorobyov} E.~I., {Skliarevskii} A.~M., {Elbakyan} V.~G., {Pavlyuchenkov} Y.,
  {Akimkin} V., {Guedel} M., 2019, Astron. and Astrophys., 627, A154

\bibitem[{{Williams} \& {Cieza}(2011)}]{2011ARA&A..49...67W}
{Williams} J.~P., {Cieza} L.~A., 2011, Annual Rev. of Astron. and Astrophys.,
  49, 67

\bibitem[{{Zhu} {et~al}\mbox{.}(2009){Zhu}, {Hartmann}, {Gammie}, \&
  {McKinney}}]{2009ApJ...701..620Z}
{Zhu} Z., {Hartmann} L., {Gammie} C., {McKinney} J.~C., 2009, \apj, 701, 620

\end{thebibliography}

\end{document}